\documentclass{emulateapj}

\usepackage{epsfig,graphicx}
\usepackage{natbib}

\shorttitle{Insights into the 300\,km\,s$^{-1}$ stellar stream near Segue\,1}
\shortauthors{Frebel et al.}

\begin{document}

\title{The 300\,km\,s$^{-1}$ stellar stream near Segue\,1: Insights From
  High-Resolution Spectroscopy of its brightest star\altaffilmark{1}}

\author{
Anna Frebel\altaffilmark{2},
Ragnhild Lunnan\altaffilmark{3},
Andrew R. Casey\altaffilmark{4,2},
John E. Norris\altaffilmark{4},
Rosemary F. G. Wyse\altaffilmark{5},
Gerard Gilmore\altaffilmark{6}}

\altaffiltext{1}{This paper includes data gathered with the 6.5 meter
  Magellan Telescopes located at Las Campanas Observatory, Chile.}

\altaffiltext{2}{Massachusetts Institute of Technology, Kavli Institute for
  Astrophysics and Space Research, 77 Massachusetts Avenue, Cambridge,
  MA 02139, USA}

\altaffiltext{3}{Harvard-Smithsonian Center for Astrophysics, 60 Garden
  Street, Cambridge, MA 02138, USA}

\altaffiltext{4}{Research School of Astronomy \& Astrophysics, The
  Australian National University, Mount Stromlo Observatory, Cotter
  Road, Weston, ACT 2611, Australia}

\altaffiltext{5}{The Johns Hopkins University, Department of Physics \&
  Astronomy, 300 N. Charles Street, Baltimore, MD 21218, USA}

\altaffiltext{6}{Institute of Astronomy, University of Cambridge, Madingley
  Road, Cambridge CB3 0HA, UK}

\begin{abstract}
We present a chemical abundance analysis of 300S-1, the brightest
likely member star of the 300\,km\,s$^{-1}$ stream near the faint
satellite galaxy Segue\,1. From a high-resolution Magellan/MIKE
spectrum we determine a metallicity of $\mbox{[Fe/H]} = -1.46 \pm 0.05
\pm 0.23$ (random and systematic uncertainties) for star 300S-1, and
find an abundance pattern similar to typical halo stars at this
metallicity. Comparing our stellar parameters to theoretical
isochrones, we estimate a distance of $18 \pm 7$\,kpc. Both the
metallicity and distance estimates are in good agreement with what can
be inferred from comparing the SDSS photometric data of the stream
stars to globular cluster sequences. While several other structures
overlap with the stream in this part of the sky, the combination of
kinematic, chemical and distance information makes it unlikely that
these stars are associated with either the Segue\,1 galaxy, the
Sagittarius stream or the Orphan stream. Streams with halo-like
abundance signatures, such as the 300\,km\,s$^{-1}$ stream, present
another observational piece for understanding the accretion history of
the Galactic halo.

\end{abstract}



\section{Introduction}
\label{sec:intro}

In the $\Lambda$CDM model, structure builds up hierarchically, with
smaller objects merging to build larger ones. One consequence of this
is stellar streams in the Galactic halo, produced as accreted objects
are tidally disrupted (see e.g. \citealt{Lynden-Bell1995} and
references therein). A well-known example is the stream extending from
the Sagittarius dwarf spheroidal, which has been traced more than one
full wrap around the Milky Way (e.g., \citealt{Ibata1994,
  Majewski2003}). In a part of the sky dubbed the ``Field of Streams''
\citep{Belokurov2006b}, at least two wraps of the Sagittarius stream
as well as several other structures are visible.

A new velocity structure in this field has recently been discovered
overlapping with the ultra-faint object Segue\,1
\citep{Belokurov2007}: Radial velocity data to determine Segue\,1's
velocity dispersion by \citet{Geha2009} revealed a group of stars
moving near 300\,km\,s$^{-1}$ (with dispersion of
$\sim10$\,km\,s$^{-1}$). For comparison, the Segue\,1 dwarf has a mean
velocity of 208\,km\,s$^{-1}$ with a dispersion of 3\,km\,s$^{-1}$
(Simon et al. 2011; hereafter S11)\nocite{Simon2011}.  The same
overdensity of stars at this velocity was seen by Norris et al. (2010;
hereafter N10)\nocite{Norris2010a}, and in the larger spectroscopic
sample of Segue\,1 stars of S11. The stars have so far been
interpreted as part of a stellar stream independent of Segue\,1, and
we shall refer to it as the ``300\,km\,s$^{-1}$ stream'' or ``300S''
throughout this paper. Since the full extent of this structure has not
been mapped out, however, it is also possible that these stars belong
to a bound object.

From SDSS photometry, one can obtain an estimate of the distance and
metallicity of the 300\,km\,s$^{-1}$ stream by comparing the
color-magnitude diagram to globular cluster sequences and
isochrones. To further characterize the stream chemically, however,
high-resolution spectroscopy is necessary. Here, we present the first
high-resolution spectrum and detailed abundance analysis of a star in
300S.

This paper is organized as follows. In Section~\ref{sec:medres}, we
briefly discuss the sample from which the main target for
high-resolution spectroscopy was selected. The observations and basic
spectral analysis of our target star is presented in
Section~\ref{sec:obs}, and the abundance analysis in
Section~\ref{sec:abund}. We interpret the results in Section~5, and
discuss the nature and potential origin of the stream in
Section~\ref{sec:conc}.

\section{Stream sample and target selection}
\label{sec:medres}

There are two existing medium-resolution spectroscopic studies of the
region around the ultra-faint dwarf galaxy Segue\,1, N10 and
S11. N10's data were obtained with the Anglo-Australian Telescope's
AAOmega spectrograph, which can take simultaneous spectra of 400
targets over a field 2$^{\circ}$ in diameter. They were targeting
stars in the RGB locus. S11, on the other hand, were using the {\small
  DEIMOS} spectrograph on the Keck\,II telescope, focusing on an area
within $\sim 15$\arcmin\, of the center of Segue\,1, and going deeper
than N10. Table~\ref{tab:members} lists all stars (52 total) in the
two samples with heliocentric radial velocities higher than
240\,km\,s$^{-1}$. This value corresponds to about the cutoff of
Segue\,1 dwarf galaxy stars in S11 (their Figure 3). It conservatively
includes all stars with velocities that are not consistent with
Segue\,1 membership, and hence potential 300S candidates. The
estimated velocity uncertainty for the N10 stars is 10\,km\,s$^{-1}$;
the individual velocity uncertainties for S11 stars are shown in the
table.  The list includes four targets from the AAOmega sample not
published in N10 due to slightly lower confidence in the radial
velocity measurement. We also list photometry from the Sloan Digital
Sky Survey, DR7 \citep{Abazajian2009}.

\begin{figure*}
 \begin{center}
  \begin{tabular}{cc}
    \includegraphics[width=8.5cm,clip=true,bbllx=70, bblly=362,bburx=545, bbury=714]{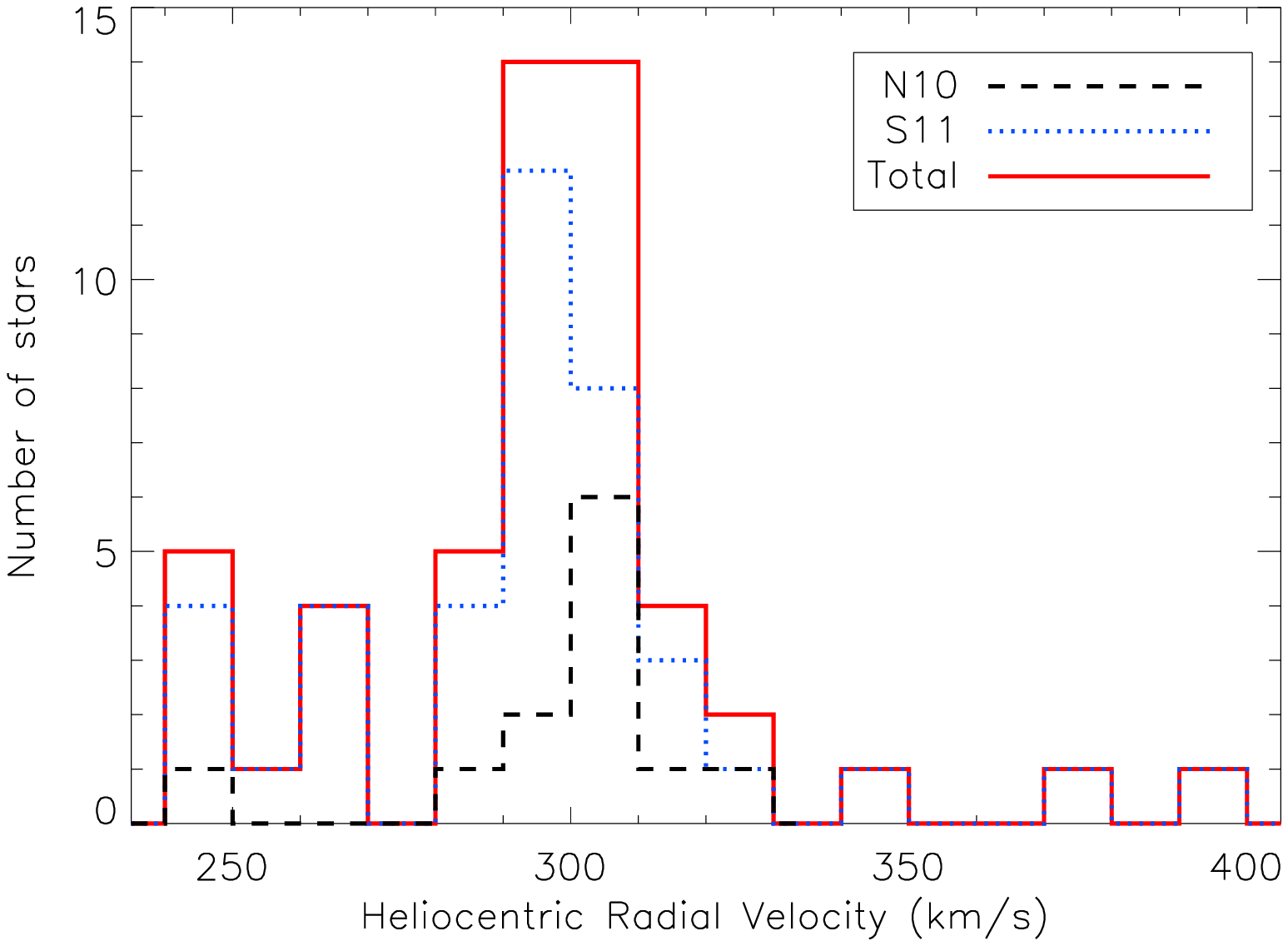} & 
    \includegraphics[width=8.5cm,clip=true,bbllx=12, bblly=178,bburx=600, bbury=613]{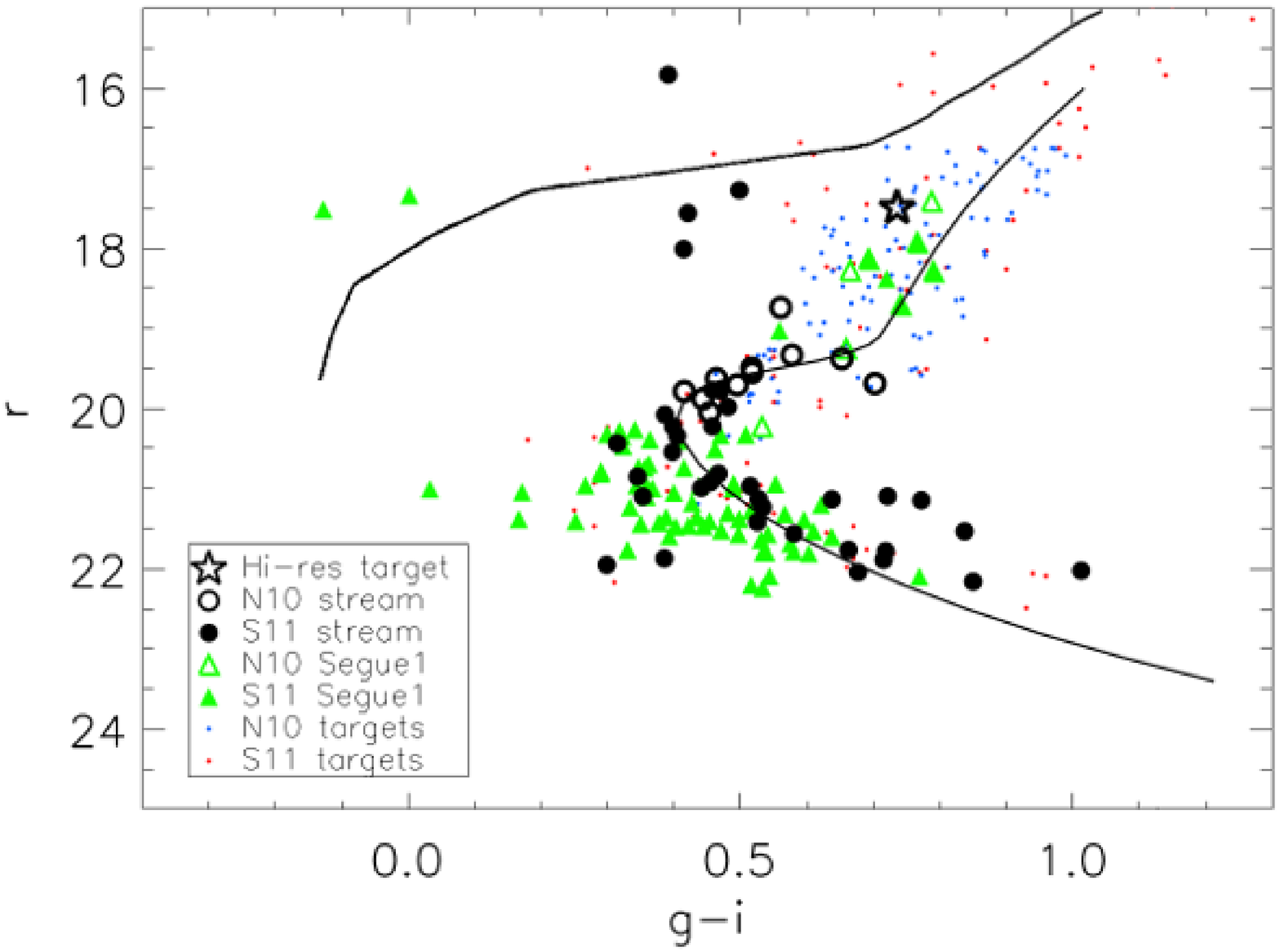} \\

    \includegraphics[width=8.5cm,clip=true,bbllx=50, bblly=360,bburx=502, bbury=714]{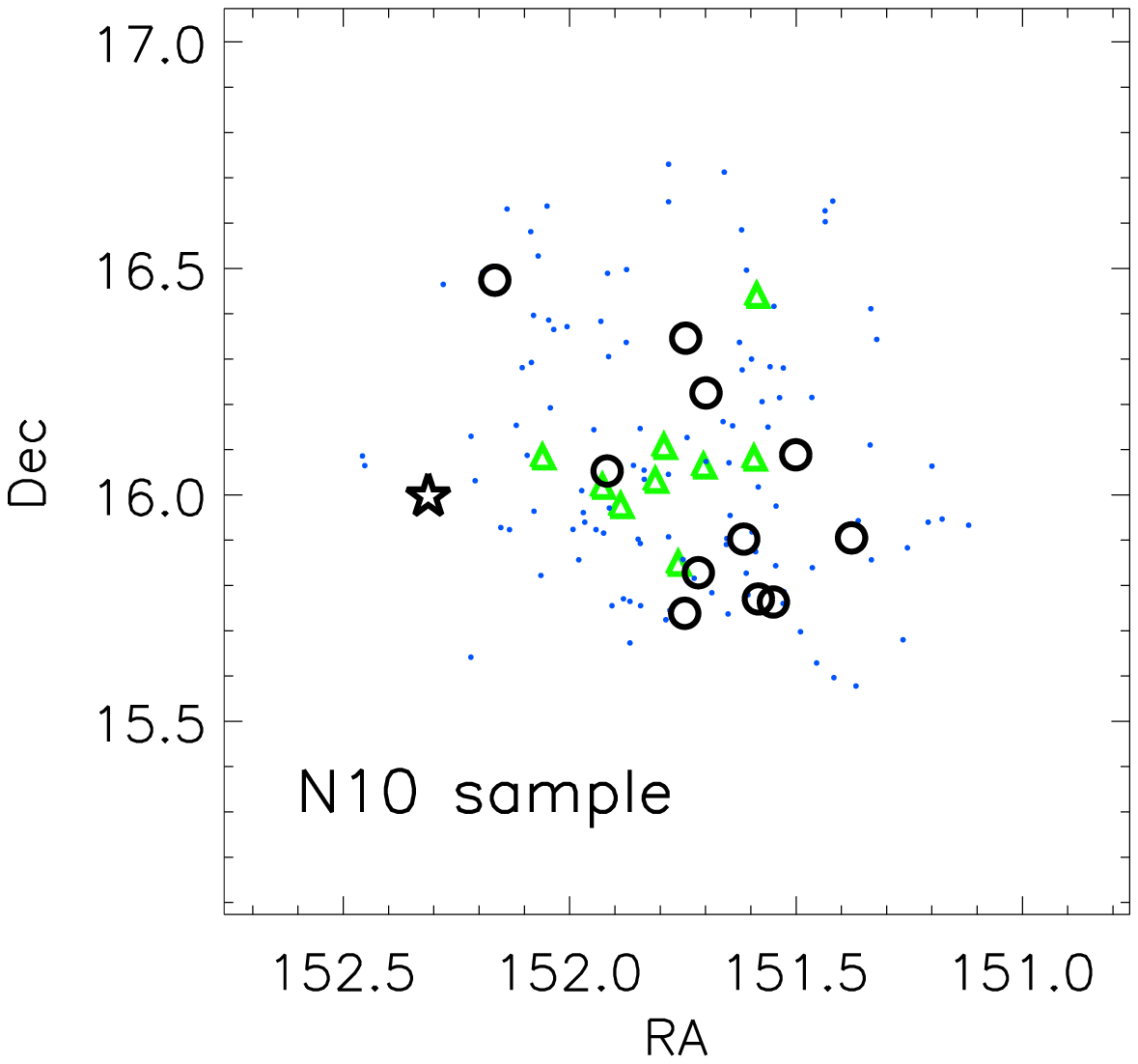}  & 
    \includegraphics[width=8.5cm,clip=true,bbllx=50, bblly=360,bburx=502, bbury=714]{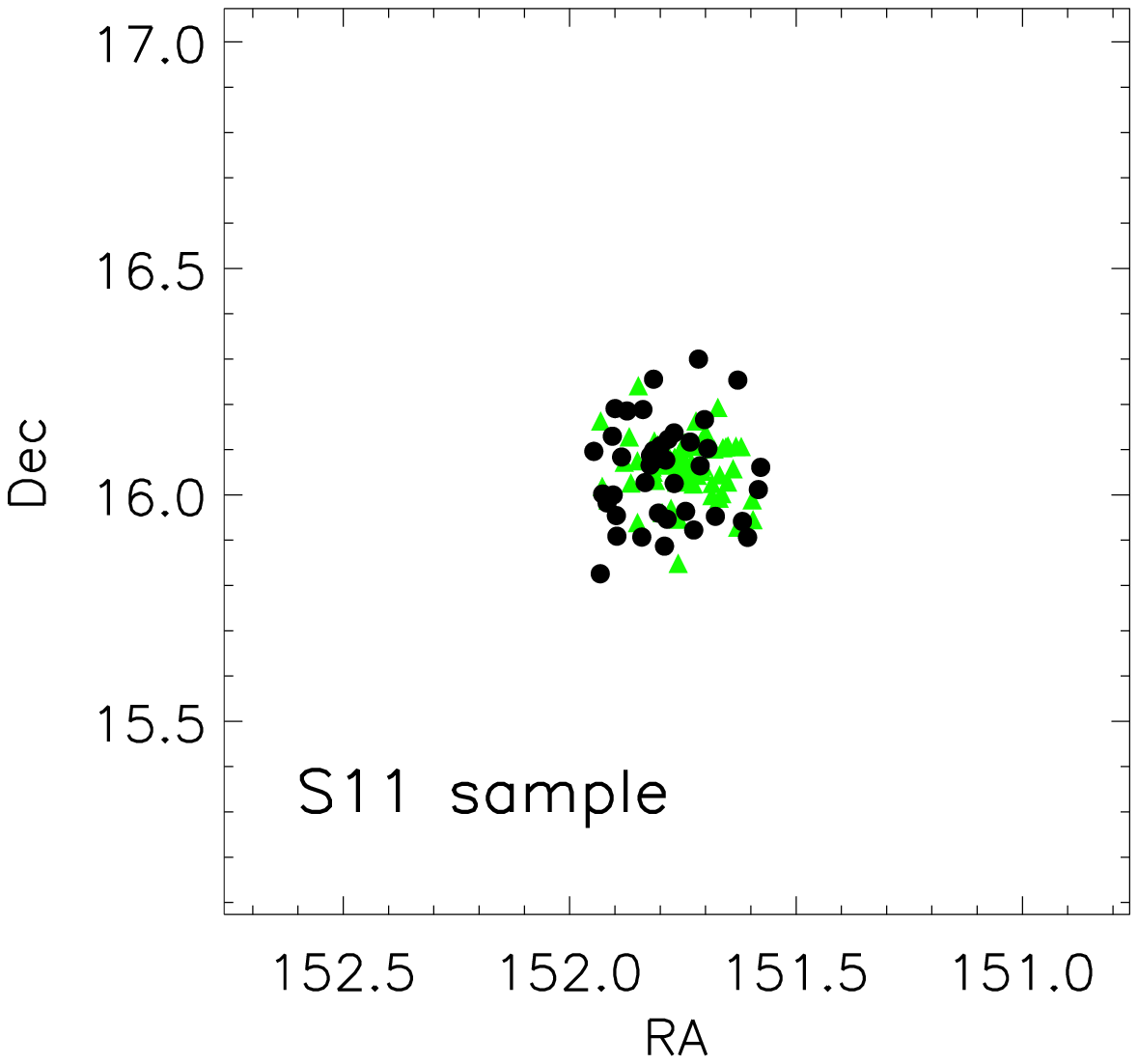}  \\

  \end{tabular}
  \caption{\scriptsize Summarizing the properties of the stream star
    candidates found in N10 and S11. Top left: Heliocentric radial
    velocity histogram of all stars in the combined N10 and S11
    samples with velocities greater than 240\,km\,s$^{-1}$. The black
    dashed line shows stars in the N10 sample, the blue dotted lines
    show stars in the S11 sample, and the solid red line shows the
    total. Top right: Color-magnitude diagram of the 300\,km\,s$^{-1}$
    stream candidates found in S11 (filled circles) and N10 (open
    circles). The open star symbol denotes our target. The green
    triangles show radial velocity members of the ultra-faint dwarf
    galaxy Segue\,1, among which the stream stars were
    discovered. Blue (N10) and red (S11) dots show the remaining stars
    that still meet the photometric cuts in either sample, but are not
    radial velocity members of Segue\,1 or the stream. Photometry is
    from the SDSS-DR7 \citep{Abazajian2009}. The black line shows the
    globular cluster sequence of M5 \citep{An2008}, shifted according
    to reddening ($E(B-V)=0.03$) and a best-fit distance of 18\,kpc. A
    rough HB is shown to guide the eye. Bottom: Coordinate plots of
    stream candidate stars, showing the N10 sample (left) and S11
    sample (right). The box size represents the 2$^{\circ}$ field of
    view covered by N10 centered on Segue\,1; S11 on the other hand
    only observed stars within 15\arcmin\, of Segue\,1's
    center. Symbols as in the color-magnitude diagram.}
  \label{fig:col_mag}
 \end{center}
\end{figure*}

Figure~\ref{fig:col_mag} summarizes the properties of the two
samples. The top left panel shows a histogram of the heliocentric
radial velocities measured. The central peak ($\sim$
270-330\,km\,s$^{-1}$) has a mean velocity of $300.4 \pm 1.1\,
\mbox{km\,s}^{-1}$ and a dispersion of $10.3 \pm
1.2\,\mbox{km\,s}^{-1}$. Considering the independent N10 and S11
samples separately we still arrive at a dispersion of 10\,km\,s$^{-1}$
for the central peak, and similar means ($303.3 \pm 3$\,km\,s$^{-1}$
for N10 versus $299.1 \pm 1$\,km\,s$^{-1}$ for S11). It is not clear
whether the intermediate ``bridge'' candidates with 240\,km\,s$^{-1}$
$< v_{helio} < 270$\,km\,s$^{-1}$ or the three extreme-velocity stars
with $v_{helio} > 340$\,km\,s$^{-1}$ should be considered part of the
same structure, but follow-up observations tracing the stream over a
larger field of view could resolve this.

The top right panel of Figure~\ref{fig:col_mag} shows the
color-magnitude diagram of all the stream candidate stars from
Table~\ref{tab:members}. Here, stream candidates are shown as black
circles. (Three stars redder than $(g-i) = 1.2 $ are not shown.) For
comparison, the green triangles show radial velocity members of the
Segue\,1 dwarf galaxy, which was the target of the N10 and S11
studies. Open symbols show stars from the N10/AAO sample, while filled
symbols are from the S11 sample. Note that the N10 sample covers a
much larger field of view but does not go as deep -- as a result, most
of the bright stream stars, including our target (star symbol), show
up in this sample. In order to illustrate the photometric criteria
used in the samples, the blue and red dots show the stars observed by
N10 and S11, respectively, that met their photometric selection cut
for follow-up spectroscopy, but are radial velocity non-members of
Segue\,1 and the stream.

The solid line shows the M5 cluster sequence \citep{An2008}, which
gives a good fit to the stream main sequence and subgiant branch when
shifted to a distance of 18\,kpc. The metallicity of M5 is
$\mbox{[Fe/H]} = -1.3$ (e.g., \citealt{Carretta2009}), and the SDSS
photometry suggests that the 300S is slightly more metal-rich than
Segue\,1 \citep{Simon2011}. We note, however, that the red giant
branch of the stream is bluer than this sequence, and in prinicple,
would be better fit by the RGB of the more metal-poor globular cluster
M92. If done so, the better populated turnoff region is not well
fitted, so we adopt the M5 sequence.

There are six stream candidate stars brighter than $r = 19$, and thus
possible candidates for high-resolution spectroscopy. Our chosen
target (SDSSJ100914.95+155948.4; the first star in
Table~\ref{tab:members}) is marked with an open star symbol. It is
identified as ``Segue\,1-11'' in N10, but we shall refer to it by
``300S-1'' throughout this paper. Its location in the color-magnitude
diagram indicates that it is most likely a red giant, though it could
also be consistent with a horizontal branch star at this distance. Both its
colors, and the radial velocity of 301\,km\,s$^{-1}$ measured by N10
are consistent with stream membership.  A medium resolution spectrum,
taken as part of the N10 campaign, indicates that 300S-1 is more
metal-rich than a typical Segue\,1 metallicity -- consistent with
S10's prediction based on isochrone fitting.  Accordingly, we deemed
it the best candidate for high-resolution spectroscopy follow-up.

The nature of the other five bright stars is less clear. As noted in
\citet{Geha2009}, random halo stars at these extreme velocities are
very rare. If they indeed were turnoff stars, the stream would have a
coherent velocity over four magnitudes in distance modulus, which
seems unlikely. The three stars around $r\sim 17.5$ could be red
horizontal branch stars, but if so, raises the question of why we do
not see more red giants when assuming the numbers of horizontal branch stars
and red giants to be roughly equal. The nature of the very bright star
at $r = 15.8$ is also not understood but additional data of these
brighter stars could provide more insight.

Finally, the spatial coverage of the two samples is shown in the two
bottom panels. The brighter stream stars in the N10 sample (left)
extend at least over 1$^{\circ}$ on the sky; the deeper sample of S11
(right) only covers a 15\arcmin\, radius around the center of
Segue\,1. The box size here represents the full field of view observed
in the N10 sample, and so for the lower left panels, again, all stars
observed by N10 are shown to better illustrate the
distribution. New photometric observations to determine the full extent of
this stream on the sky, as well as deeper observations in a larger
region than that covered by S11, would be very important for better
understanding and characterizing this structure.

\begin{deluxetable*}{lcccccccc}
\tabletypesize{\scriptsize}
\tablecaption{Stream Member Candidate Stars\label{tab:members}}
\tablehead{
  \colhead{Identifier} &
  \colhead{R. A.} &
  \colhead{Dec.} &
  \colhead{$g$} &
  \colhead{$r$} &
  \colhead{$i$} &
  \colhead{$g \-- r$} &
  \colhead{$V_{helio}$} &
  \colhead{Ref} \\
 & (J2000) & (J2000) & & & & & (km s$^{-1}$) &  
}
\startdata
300S-1\tablenotemark{a} & 10 09 15.0 & $+$15 59 48.4 &  17.99 & 17.49 & 17.26 & 0.73 & 307 & N10 \\
300S-2 & 10 07 40.1 & $+$16 03 09.7 &  19.96 & 19.55 & 19.44 & 0.52 & 298/303.1$\pm$3.1 & N10/S11    \\
300S-3 & 10 06 59.0 & $+$15 44 18.8 &  19.72 & 19.33 & 19.14 & 0.58 & 307 & N10     \\
300S-4 & 10 06 51.8 & $+$15 49 41.5 &  20.36 & 20.04 & 19.91 & 0.45 & 300 & AAO     \\
300S-5 & 10 06 20.0 & $+$15 46 12.7 &  20.09 & 19.78 & 19.67 & 0.42 & 315 & AAO     \\
300S-6 & 10 06 12.0 & $+$15 45 48.4 &  20.08 & 19.62 & 19.61 & 0.47 & 327 & N10     \\
300S-7 & 10 05 30.6 & $+$15 54 18.1 &  19.83 & 19.49 & 19.31 & 0.52 & 305 & N10     \\
300S-8 & 10 06 27.7 & $+$15 54 08.6 &  20.15 & 19.88 & 19.71 & 0.44 & 300 & AAO     \\
300S-9 & 10 06 47.7 & $+$16 13 30.1 &  20.01 & 19.70 & 19.51 & 0.50 & 300 & AAO     \\
300S-10 & 10 06 58.5 & $+$16 20 45.6 &  19.84 & 19.37 & 19.18 & 0.66 & 295 & N10     \\
300S-11 & 10 08 39.7 & $+$16 28 26.7 &  19.16 & 18.73 & 18.60 & 0.56 & 288 & N10     \\
300S-12 & 10 06 00.2 & $+$16 05 18.6 &  20.22 & 19.68 & 19.51 & 0.71 & 242 & N10      \\
300S-13 & 10 06 18.7 & $+$16 03 39.0 &  21.67 & 21.14 & 20.90 & 0.77 & 268.6  $\pm$ 3.1 & S11     \\
300S-14 & 10 06 20.0 & $+$16 00 42.1 &  21.70 & 21.41 & 21.18 & 0.52 & 305.5  $\pm$ 4.3 & S11     \\
300S-15 & 10 06 30.9 & $+$16 15 12.1 &  21.47 & 21.13 & 20.95 & 0.52 & 321.0  $\pm$ 8.8 & S11     \\
300S-16 & 10 06 42.8 & $+$15 57 09.3 &  21.56 & 21.23 & 21.02 & 0.54 & 291.9  $\pm$ 6.4 & S11     \\
300S-17 & 10 06 46.8 & $+$16 06 08.3 &  20.53 & 20.22 & 20.07 & 0.46 & 294.5  $\pm$ 2.9 & S11     \\
300S-18 & 10 06 48.5 & $+$16 09 58.1 &  20.27 & 19.98 & 19.79 & 0.48 & 306.0  $\pm$ 2.6 & S11     \\
300S-19 & 10 06 50.8 & $+$16 03 51.2 &  22.13 & 21.95 & 21.83 & 0.30 & 312.6  $\pm$11.9 & G09, S11     \\
300S-20 & 10 06 54.2 & $+$15 55 20.7 &  22.12 & 21.53 & 21.29 & 0.83 & 268.5  $\pm$ 6.2 & S11     \\
300S-21 & 10 06 56.1 & $+$16 06 60.0 &  21.34 & 21.09 & 20.99 & 0.35 & 302.0  $\pm$ 2.6 & S11     \\
300S-22 & 10 06 58.5 & $+$15 57 48.9 &  21.56 & 21.09 & 20.84 & 0.72 & 290.5  $\pm$ 3.6 & S11     \\
300S-23 & 10 07 04.6 & $+$16 01 30.8 &  21.22 & 20.81 & 20.76 & 0.46 & 295.8  $\pm$ 3.9 & S11     \\
300S-24 & 10 07 04.6 & $+$16 08 12.6 &  21.08 & 20.85 & 20.74 & 0.34 & 296.9  $\pm$ 3.4 & S11     \\
300S-25 & 10 07 08.4 & $+$15 56 46.3 &  22.03 & 21.56 & 21.45 & 0.58 & 286.3  $\pm$ 5.4 & S11     \\
300S-26 & 10 07 09.1 & $+$16 04 36.6 &  22.29 & 21.88 & 21.57 & 0.72 & 312.7  $\pm$ 6.4 & S11     \\
300S-27 & 10 07 09.7 & $+$15 53 12.3 &  16.11 & 15.83 & 15.72 & 0.39 & 303.4  $\pm$ 2.2 & S11     \\
300S-28 & 10 07 13.0 & $+$15 57 34.8 &  18.28 & 18.00 & 17.87 & 0.41 & 307.9  $\pm$ 2.4 & S11     \\
300S-29 & 10 07 13.7 & $+$16 04 44.8 &  22.13 & 21.76 & 21.47 & 0.66 & 293.4  $\pm$ 4.8 & G09, S11     \\
300S-30 & 10 07 15.5 & $+$16 05 52.1 &  20.36 & 20.07 & 19.97 & 0.39 & 282.0  $\pm$ 2.8 & S11     \\
300S-31 & 10 07 15.5 & $+$16 15 19.1 &  20.74 & 20.43 & 20.42 & 0.32 & 266.8  $\pm$ 3.1 & S11     \\
300S-32 & 10 07 17.2 & $+$16 05 11.9 &  22.10 & 21.78 & 21.38 & 0.72 & 266.3  $\pm$ 4.4 & S11     \\
300S-33 & 10 07 17.4 & $+$16 03 55.6 &  20.11 & 19.77 & 19.64 & 0.47 & 295.9  $\pm$ 2.4 & G09, S11    \\
300S-34 & 10 07 20.0 & $+$16 01 37.5 &  17.62 & 17.27 & 17.12 & 0.50 & 312.7  $\pm$ 2.2 & S11     \\
300S-35 & 10 07 21.2 & $+$16 11 18.2 &  20.99 & 19.73 & 19.20 & 1.79 & 281.6  $\pm$ 2.4 & S11     \\
300S-36 & 10 07 21.8 & $+$15 54 24.5 &  20.61 & 20.22 & 20.22 & 0.39 & 307.2  $\pm$ 5.5 & S11     \\
300S-37 & 10 07 29.6 & $+$16 11 07.1 &  20.35 & 19.34 & 19.02 & 1.33 & 309.6  $\pm$ 2.2 & S11     \\
300S-38 & 10 07 32.5 & $+$16 05 00.5 &  22.58 & 22.04 & 21.91 & 0.67 & 281.1  $\pm$ 6.9 & G09, S11     \\
300S-39 & 10 07 35.0 & $+$15 54 31.5 &  20.78 & 20.54 & 20.39 & 0.39 & 303.2  $\pm$ 2.8 & S11     \\
300S-40 & 10 07 37.3 & $+$16 07 46.2 &  21.25 & 20.99 & 20.81 & 0.44 & 296.0  $\pm$ 3.9 & S11     \\
300S-41 & 10 07 40.2 & $+$15 58 55.6 &  21.32 & 20.96 & 20.80 & 0.52 & 295.8  $\pm$ 3.8 & S11     \\
300S-42 & 10 07 42.5 & $+$16 00 06.8 &  22.47 & 22.02 & 21.46 & 1.01 & 296.6  $\pm$10.3 & S11     \\
300S-43 & 10 07 43.8 & $+$15 49 32.9 &  22.47 & 20.98 & 20.36 & 2.11 & 299.2  $\pm$ 2.5 & S11     \\
300S-44 & 10 07 47.2 & $+$16 05 45.5 &  20.13 & 19.77 & 19.67 & 0.46 & 294.3  $\pm$ 2.4 & S11     \\
300S-45 & 10 07 35.9 & $+$16 11 25.7 &  23.79 & 22.08 & 22.00 & 1.79 & 242.2  $\pm$ 7.8 & S11     \\
300S-46 & 10 06 25.7 & $+$15 54 22.1 &  21.58 & 21.13 & 20.95 & 0.63 & 244.3  $\pm$ 5.6 & S11     \\
300S-47 & 10 07 11.8 & $+$16 06 30.4 &  22.80 & 22.16 & 21.95 & 0.85 & 247.1  $\pm$15.9 & S11     \\
300S-48 & 10 07 07.8 & $+$16 07 21.5 &  20.64 & 20.34 & 20.23 & 0.41 & 247.7  $\pm$ 2.8 & S11     \\
300S-49 & 10 07 35.2 & $+$15 57 15.3 &  22.53 & 21.13 & 20.41 & 2.12 & 255.1  $\pm$ 3.0 & S11     \\
300S-50 & 10 06 28.4 & $+$15 56 28.8 &  17.85 & 17.56 & 17.43 & 0.42 & 347.1  $\pm$ 2.9 & S11     \\
300S-51 & 10 07 36.9 & $+$15 59 58.9 &  21.99 & 21.87 & 21.60 & 0.39 & 373.0  $\pm$ 6.0 & S11     \\
300S-52 & 10 06 51.7 & $+$16 17 59.2 &  21.40 & 20.92 & 20.95 & 0.45 & 394.9  $\pm$ 8.9 & S11 
\enddata
\tablenotetext{a}{Target star, referenced as Segue1-11 in N10}
\end{deluxetable*}

\section{Observations and Data Analysis}
\label{sec:obs}

\subsection{Observations}

We obtained a spectrum of 300S-1 with the MIKE spectrograph \citep{mike}
on the Magellan Clay telescope in March 2010. The total exposure time for
this $V = 17.6$\,mag star was 4.5 hours, distributed over six exposures to
 allow for removal of cosmic rays. MIKE spectra have
nearly full optical wavelength coverage from $\sim3500$-9000\,{\AA}. Using
a 1.0\arcsec\, slit and $2\times2$ on-chip binning, a resolution of
$\sim22,000$ is achieved in the red, and $\sim28,000$ in the blue
wavelength regime. 

The data were reduced using an echelle data reduction pipeline made
for MIKE\footnote{Available at
  http://obs.carnegiescience.edu/Code/python}.  The reduced individual
orders were normalized and merged to produce final one-dimensional
blue and red spectra ready for the analysis. The $S/N$ of the data of
this faint object is modest: 17 at $\sim4500$\,{\AA} and 20 at
$\sim5200$\,{\AA}.

\begin{figure}
 \begin{center}
  \includegraphics[clip=true,width=8.5cm,bbllx=55, bblly=12,
   bburx=510, bbury=370]{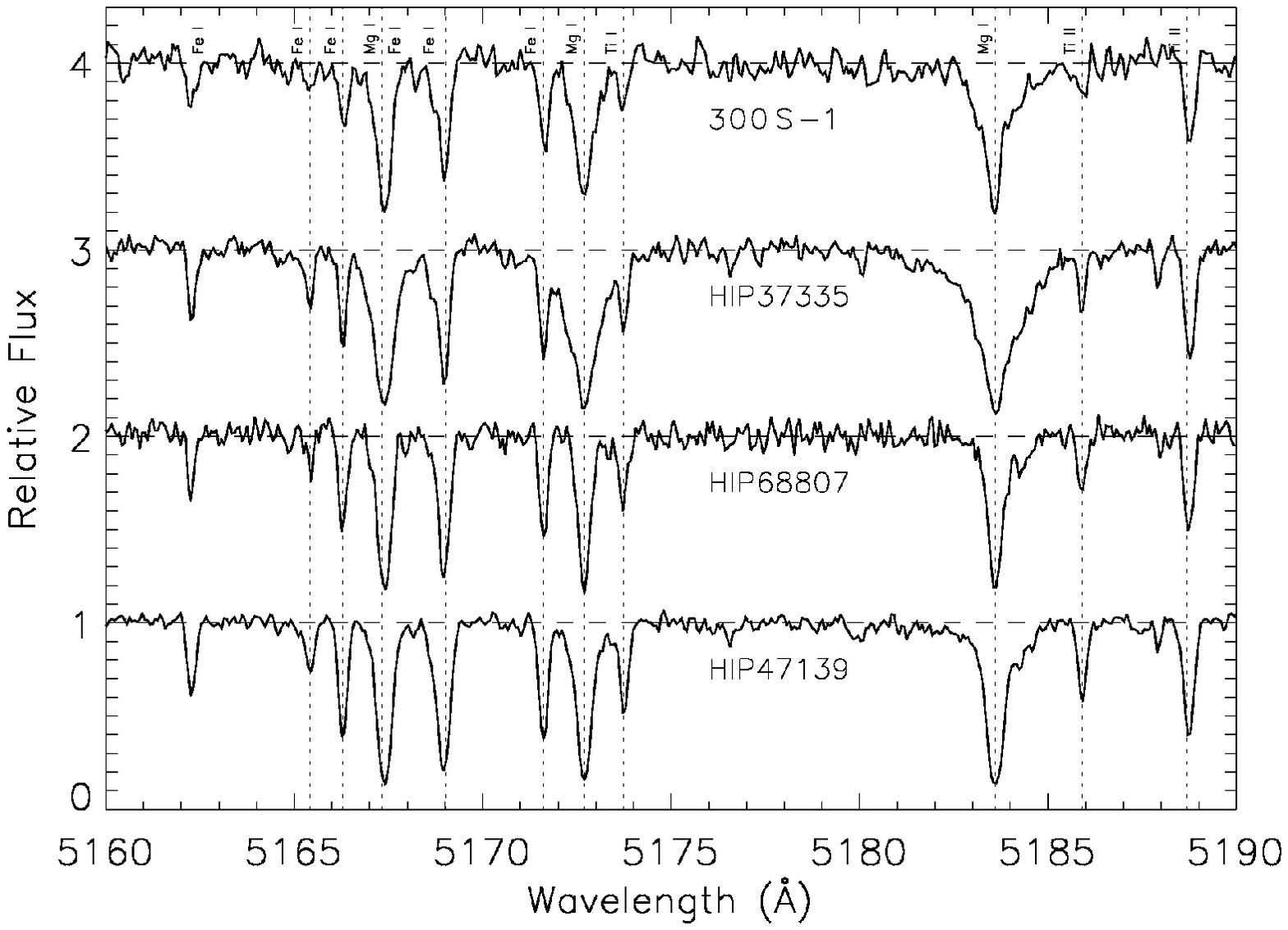}
  \caption{High-resolution spectrum of our target star, 300S-1, and
    the three comparison stars from the F00 sample, in the region
    around the Mg b lines. Shown here are the short exposures of the
    comparison stars, taken to obtain a similar resolution and $S/N$ as
    the 300S-1 spectrum.}
  \label{fig:mglines}
 \end{center}
\end{figure}

In addition to our main target, we also took MIKE spectra of three
bright comparison stars chosen from the Fulbright (2000; hereafter
F00)\nocite{Fulbright2000} sample in March 2011. These comparison
stars were chosen based on having stellar parameters and metallicities
that bracketed our first estimate for 300S-1. Table~\ref{tab:targets}
summarizes the observations of our targets.  Spectra were taken with a
1.0\arcsec\, slit and short exposures, in order to get a similar data
quality and $S/N$ to that of the 300S-1
spectrum. Figure~\ref{fig:mglines} shows the spectrum of 300S-1 and
the short-exposure spectra of the comparison stars in the region
around the Mg b lines at 5170\,{\AA}. By comparing stellar parameters
and abundances derived from these short-exposure spectra to the
published values, we are able to assess the accuracy of our low $S/N$
spectrum of 300S-1.

\begin{deluxetable*}{lcccccccc}
\tabletypesize{\scriptsize}
\tablecaption{Observed Targets \label{tab:targets}}
\tablehead{
  \colhead{Name} &
  \colhead{R. A.} &
  \colhead{Dec.} &
  \colhead{$V$} &
  \colhead{$B \-- V$} & 
  \colhead{UT Date} &
  \colhead{UT Start} &
  \colhead{$t_{exp}$} \\
 & (J2000) & (J2000) & & & & & ($s$)   
}
\startdata  
300S-1     &  10 09 15.0 &  $+$15 59 48.4 & 17.70\tablenotemark{a} &  0.68\tablenotemark{a} & 03/08/2010 & 02:35 & 3000 \\
& & & & & 03/08/2010 & 05:30 & 3600 \\
& & & & & 03/19/2010 & 05:06 & 3000 \\
& & & & & 03/22/2010 & 00:37 & 3100 \\
& & & & & 03/22/2010 & 05:22 & 2800 \\
& & & & & 03/23/2010 & 05:08 & 700 \\
HIP37335   &  07 39 50.1 &  $-$01 31 20.4 & 9.25 &  0.82 & 03/13/2011 & 23:41 & 7      \\
HIP47139   &  09 36 20.0 &  $-$20 53 14.8 & 8.34 &  1.01 & 03/12/2011 & 10:08 & 1      \\
HIP68807   &  14 05 13.0 &  $-$14 51 25.5 & 7.25 &  0.93 & 03/13/2011 & 23:53 & 5      
\enddata
\tablenotetext{a}{Transformed from SDSS photometry following \citet{Jordi2006}}
\end{deluxetable*}

\subsection{Line Strength Measurements}
\label{sec:linemes}
We obtained a first estimate for the radial velocity from the two
strong Mg b lines in the green region of the spectrum, and two
additional Mg lines in the blue. Equivalent widths were then measured
by fitting Gaussian profiles to the metal absorption lines, and our
estimate was corrected based on the mean radial velocity from all the
lines measured. Using these 366 lines, we find a heliocentric radial
velocity of $307.6$\,km\,s$^{-1}$, with a standard error of the mean
of 0.1\,km\,s$^{-1}$. This is slightly higher than the estimate of
$301$\,km\,s$^{-1}$ from N10 based on the medium resolution spectrum,
but consistent within their estimated velocity uncertainty of
10\,km\,s$^{-1}$. However, our measurement is well within the
estimated range for the 300\,km\,s$^{-1}$ stream.

The linelist used for the abundance analysis is based on lines
presented in \citet{Roederer2010}, \citet{Aoki2007b}, and
\citet{Cayrel2004}. In the instances where the same line was included
in more than one (original) linelist, the most up to date oscillator
strength was used, following Roederer et al. (2010).  This linelist
was initally compiled for work on stars more metal-poor than the
target, but besides the strongest lines, we found it to work well for
this metallicity range also.

For atomic lines, equivalent widths were measured by fitting Gaussian
profiles. Lines with reduced equivalent widths $\log$(EW/$\lambda$) $>
-4.5$ were not used for abundance determination, since they fall near
the flat part of the curve-of-growth. Given the noise in the spectra,
most lines with EW $\lesssim 20$\,m{\AA}, were cautiously excluded
from the analysis, except for when the $S/N$ in the respective
wavelength range allowed for a $>3\sigma$ detection (e.g. in the red
spectral region).

For molecular features and elements with hyperfine splitting, we used
a spectrum synthesis approach. The abundance was then determined by
matching synthesized spectra of different abundances to the observed
spectra. See Section~\ref{sec:abund} for details.

\subsection{Stellar Parameters}
\label{sec:stpar}
The stellar parameters were determined by using the iron lines in each
spectrum, by an iterative process. First, the microturbulence is fixed
by demanding that the line abundances show no trend with reduced
equivalent width ($\log \mbox{EW}/\lambda$). Similarly, effective
temperature is set by requiring no trend of abundance with excitation
potential of the lines. Finally, the gravity is fixed by requiring
that the abundance derived from Fe\,II lines agree with that obtained
from Fe\,I to within 0.05\,dex. By varying the temperature and
microturbulence, and comparing the slope to the scatter in the data,
we adopt an uncertainty of $\pm$ 150\,K in temperature, and $\pm$
0.3\,km\,s$^{-1}$ in microturbulence. Similarly, we obtain an
uncertainty in the gravity of $\pm$ 0.4\,dex by seeing how much the
gravity can be changed with Fe\,I and Fe\,II still being consistent
within their uncertainties.

Table~\ref{tab:comp_para} shows the resulting stellar parameters for
300S-1 and the three comparison stars obtained with this method, with
the values obtained by F00 in parenthesis for comparison. For HIP68807
and HIP47139, our solutions agree well with the parameters published
by F00. For HIP37335, we arrive at a slightly higher temperature and
gravity, and thus metallicity, than F00.

\begin{figure}
 \begin{center}
  \includegraphics[width=8.5cm,clip=true,bbllx=12, bblly=182,bburx=600, bbury=610]{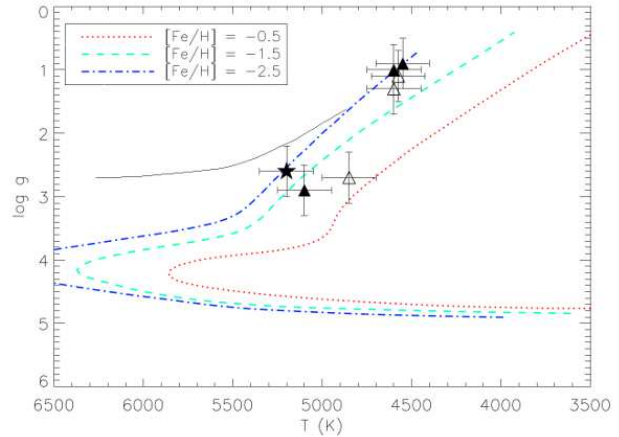}
  \caption{Adopted stellar parameters for 300S-1 (filled star), as
    well as for the three comparison stars (filled triangles). The
    open triangles show the values for the comparison stars from
    Fulbright (2000). Error bars are $\pm$ 150\,K and 0.4\,dex in
    $\log g$. Also shown are theoretical 10-Gyr isochrones at
    $\mbox{[Fe/H]} = -2.5$, $-1.5$ and $-0.5$ respectively
    \citep{Kim2002}. A metal-poor horizontal branch has been added to
    guide the eye.}
  \label{fig:isochr}
 \end{center}
\end{figure}

Figure~\ref{fig:isochr} shows the adopted stellar parameters
overplotted with theoretical 10\,Gyr isochrones \citep{Kim2002}.  Our
values agree reasonably well with the tracks within their
uncertainties.  As its position on the color-magnitude diagram
suggested, spectroscopically derived stellar parameters confirm that
300S-1 is located on the red giant branch. We note that the choice for
the age of the isochrone does not influence any conclusion since the
giant branches are nearly identical for 10 and e.g. 12\,Gyr.

For comparison, we also use the SDSS colors of 300S-1 to determine the
temperature photometrically by interpolating the SDSS {\it ugriz}
colors to the isochrones from \citet{Kim2002}, using the color tables
of Castelli (http://wwwuser.oat.ts.astro.it/castelli/).  This results
in a slightly warmer temperature (depending on which colors we use),
with T$_{\mbox{\small{eff}}} \sim 5400$\,K and $\log g \sim 3.5$. If
we instead use this set of stellar parameters, we would arrive at
[Fe/H] of $-1.3$. This is, however, well within our estimate of
uncertainty for the metallicity (see Section~\ref{sec:unc}). For the
rest of the analysis, we use the parameters derived from spectroscopy
to facilitate the relative analysis with the \citet{Fulbright2000}
stars.

\begin{deluxetable}{lcccccccc}
\tabletypesize{\scriptsize}
\tablecaption{Derived Stellar Parameters \label{tab:comp_para}}
\tablehead{
  \colhead{Name} &
  \colhead{$T_{eff}$} &
  \colhead{$\log{g}$} &
  \colhead{$[\mbox{Fe/H}]$} &
  \colhead{$v_t$} \\
 & (K) & (dex) & (dex) & (km\,s$^{-1}$)
}
\startdata  
300S-1     &  5200          &  2.6      & $-1.4$            & 1.5                \\
HIP37335\tablenotemark{a}   &  5100 (4850)   &  2.9 (2.7)& $-1.0$ ($-1.2$)   & 1.5 (1.5)           \\
HIP47139   &  4550 (4600)   &  0.9 (1.3)& $-1.6$ ($-1.4$)   & 2.3 (1.8)          \\
HIP68807   &  4600 (4575)   &  1.0 (1.1)& $-1.8$ ($-1.8$)   & 2.0 (1.9)         
\enddata

\tablenotetext{a}{While our solution for HIP37335 is warmer than what
  was found in F00, we note that it agrees with other
  literature sources for this star (e.g., \citealt{Soubiran2008,
    Cenarro2007, Peterson1981}).
\tablenotetext{}{The values in parenthesis are those determined by F00, for comparison.}}
\end{deluxetable}

\subsection{Model Atmospheres}
Our abundance analysis utilizes one-dimensional plane-parallel Kurucz
model atmospheres with overshooting and $\alpha$-enhancement
\citep{kurucz}. They are computed under the assumption of local
thermodynamic equilibrium (LTE). We use the 2010 version of the MOOG
synthesis code (first described in Sneden 1973)\nocite{moog} for this
analysis.  In this version, scattering is currently treated as true
absorption, which may have consequences for abundances derived from
lines in the blue region of the spectrum. \citet{hollek} tested how
the stellar parameters are influenced by that effect and found that
temperature and gravities, and hence [Fe/H] are somewhat lower (0.1 to
0.2\,dex) when scattering is properly treated. This average abundance
difference could be explained by the fact that abundances of metal
lines at lower wavelengths (below $\sim4200$\,{\AA}) yield lower
abundances when the scattering is treated as Rayleigh scattering. They
used metal lines down to 3750\,{\AA}. However, [X/Fe] values were
found to not change beyond $\sim0.05$\,dex. Similar results were found
by \citet{Frebel2010a} and \citet{venn12}. Hence, our star might be a
little more metal-poor (perhaps 0.1\,dex), because we have no
metal lines bluer than 4000\,{\AA} and the abundance ratios would not
be significantly affected by these different treatments. As discussed
below, these effects are well accounted for in our error budget, and
moreover, do not affect our conclusions regarding the nature of
300S-1.


\section{Abundance Analysis of 300S-1}
\label{sec:abund}

The derived stellar abundances of 300S-1 are summarized in
Table~\ref{tab:seg11_abund}. The uncertainties quoted are the standard
error of the mean, but we adopt a minimum uncertainty of
0.05\,dex. Solar abundances are taken from \citet{Asplund2009}.  This
section discusses the measurement and uncertainties of the different
elements in more detail.

\begin{deluxetable}{lcccccccc}
\tabletypesize{\scriptsize}
\tablecaption{300S-1 Abundances \label{tab:seg11_abund}}
\tablehead{
  \colhead{Element} &
  \colhead{$\log\epsilon (\mbox{X}_{\odot})$} &
  \colhead{$\log\epsilon (\mbox{X})$} &
  \colhead{$\sigma$} &
  \colhead{$N$} & 
  [X/H] &
  [X/Fe] \\
  & (dex) & (dex) & (dex) & & (dex) & (dex)
}
\startdata  
C (CH) & 8.43 &  7.24 & 0.21 & 2   & $-1.19$ & $+0.27 $ \\
Na\,I  & 6.24 &  4.93 & 0.08 & 3   & $-1.31$ & $+0.15 $ \\
Mg\,I  & 7.60 &  6.30 & 0.07 & 6   & $-1.30$ & $+0.16 $ \\
Al\,I  & 6.45 &$<5.00$&\nodata&2   & $<-1.45$& $<0.01 $ \\
Ca\,I  & 6.34 &  5.30 & 0.06 & 16  & $-1.04$ & $+0.42 $ \\
Sc\,II & 3.15 &  1.46 & 0.05 & 3   & $-1.69$ & $-0.23 $ \\ 
Ti\,I  & 4.95 &  3.69 & 0.07 & 18  & $-1.26$ & $+0.20 $ \\
Ti\,II & 4.95 &  3.77 & 0.06 & 17  & $-1.18$ & $+0.28 $ \\
Cr\,I  & 5.64 &  3.96 & 0.05 & 8   & $-1.68$ & $-0.22 $ \\
Mn\,I  & 5.43 &  3.41 & 0.07 & 3   & $-2.02$ & $-0.56 $ \\ 
Fe\,I  & 7.50 &  6.04 & 0.05 & 103 & $-1.46$ &   0.00   \\
Fe\,II & 7.50 &  6.08 & 0.05 & 10  & $-1.42$ &  +0.04   \\
Co\,I  & 4.99 &  3.29 & 0.12 & 3   & $-1.70$ & $-0.24$ \\ 
Ni\,I  & 6.22 &  4.73 & 0.06 & 14  & $-1.49$ & $-0.03$ \\
Zn\,I  & 4.56 &  3.18 & 0.25 & 2   & $-1.38$ & $+0.08$ \\
 
Sr\,II & 2.87 &  0.50:& 0.40 & 1   & $-2.37$:& $-0.91$: \\ 
Ba\,II & 2.18 &  0.83 & 0.21 & 2   & $-1.35$ & $+0.11$ \\
La\,II & 1.10 & $-0.38$ & 0.30 &1  & $-1.48$ & $-0.02$ \\
Eu\,II & 0.52 & $-0.39$:& 0.40 &1  & $-0.91$:& $+0.55$:
\enddata
\end{deluxetable}

\subsection{Carbon}
\begin{figure}
 \begin{center}
  \includegraphics[width=8.5cm,clip=true,bbllx=80, bblly=362,bburx=550, bbury=705]{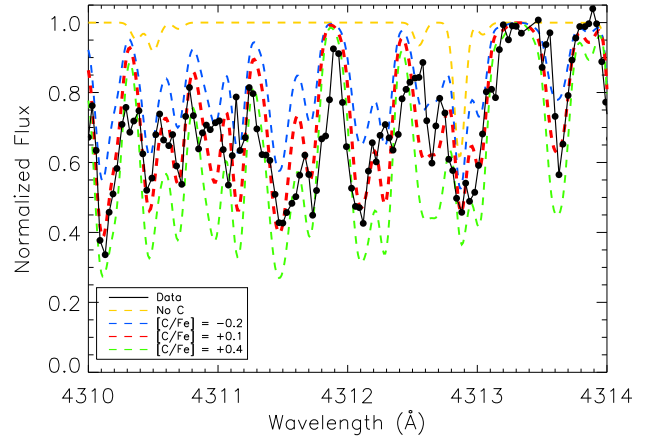}
  \caption{Example of determining carbon abundance by synthesis: The
    black dotted line shows the actual spectrum of 300S-1 at the
    carbon G-band, while the colored lines show synthesized spectra at
    different carbon abundances. }
  \label{fig:gband}
 \end{center}
\end{figure}

The carbon abundance was determined by synthesis of the carbon G-band
head at 4313\,{\AA}, and the CH feature at 4323\,{\AA}. An example of
this, comparing the observed spectrum to four synthesized spectra, is
shown in Figure~\ref{fig:gband}. Here, the thick red line shows the
carbon abundance adopted for this region,while the blue and green show
the synthesis with $\Delta \mbox{[C/Fe]} \pm 0.3$\,dex. Synthesis of
the feature at 4323\,{\AA} was done independently; the carbon
abundance quoted in Table~\ref{tab:seg11_abund} is the mean of the
two. Given the noise in the data, we adopt an uncertainty of $\pm$
0.3\,dex for each measurement.

\subsection{Light elements}
Abundances of elements without hyperfine structure were determined
from the equivalent width measurements, as described in
Section~\ref{sec:linemes}. In that case, the uncertainties listed in
Table~\ref{tab:seg11_abund} are the standard error of the mean of the
abundances determined from the individual lines for each
element. Abundances of elements with hyperfine structure (Mn, Co) were
determined by synthesis of individual lines.

In general, the abundance patterns derived from the high-resolution
spectrum are similar to those of outer halo stars at this metallicity
(also see Section~\ref{sec:ab_rat} and Figure~\ref{fig:light_el}). The
possible exception is Mg, which at $\mbox{[Mg/Fe]}=0.14$ is low
compared to the other $\alpha$-elements. We note, however, that the
derived Mg abundance is very sensitive to the assumed surface gravity
in the model, and that our Mg measurements for the three comparison
stars are lower than that measured in F00 (also see
Section~\ref{sec:unc} and Table~\ref{tab:hipabund}), so this could be
a systematic effect. Taking the $\alpha$-element abundance as (Ca + Mg
+ Ti)/3, we find $\mbox{[$\alpha$/Fe]}= +0.26 $. 300S-1 is at the low
end of $\alpha$-enhancement compared to most halo stars at this
metallicity, but still higher than $\alpha$-abundances seen in
classical dwarf spheroidal galaxies (e.g., \citealt{Tolstoy2009}).

\subsection{Neutron-capture elements}

\begin{figure*}
 \begin{center}
  \begin{tabular}{cc}
  \includegraphics[width=8.5cm,clip=true,bbllx=80, bblly=362,bburx=558, bbury=705]{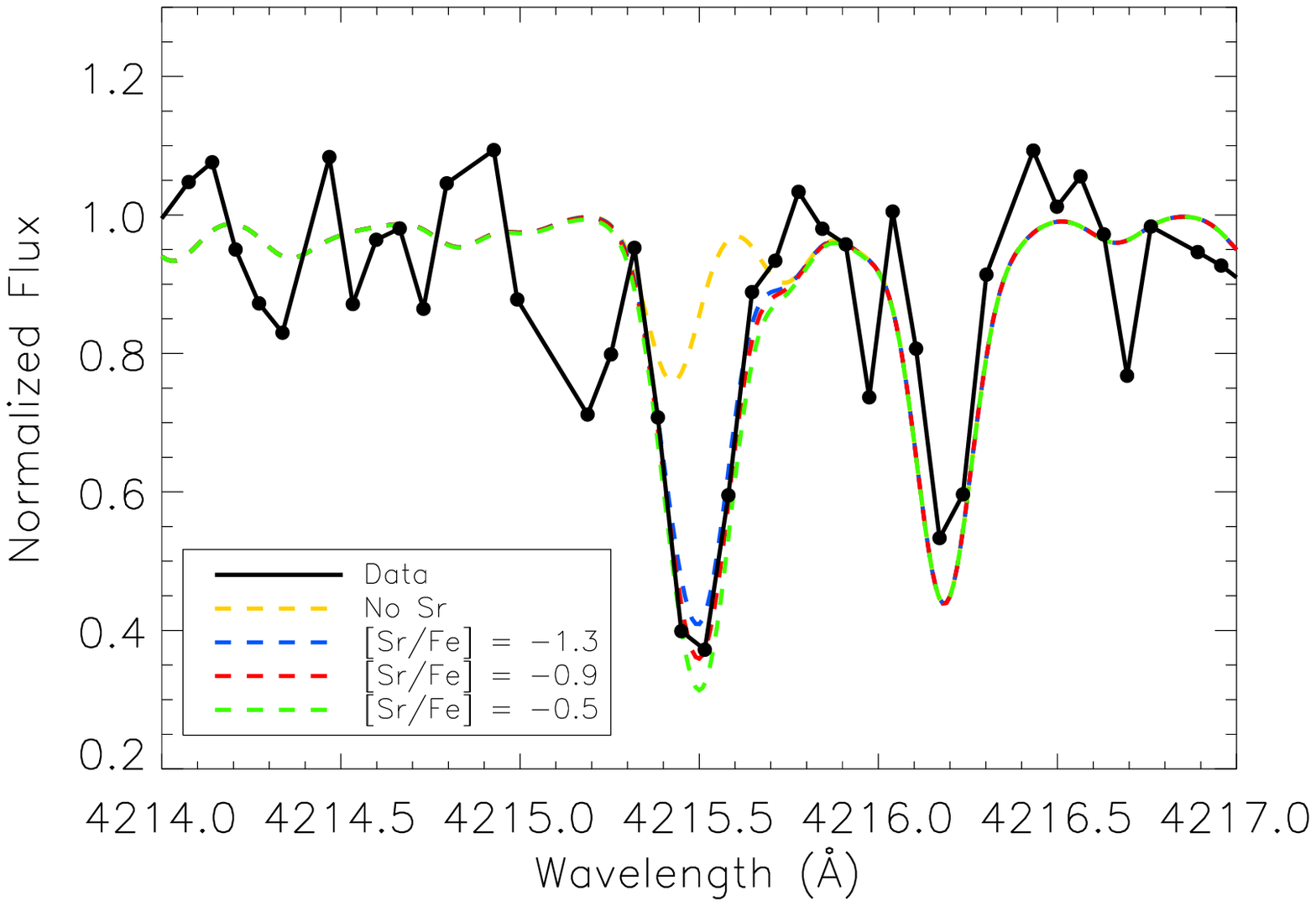} & 
  \includegraphics[width=8.5cm,clip=true,bbllx=80, bblly=362,bburx=550, bbury=705]{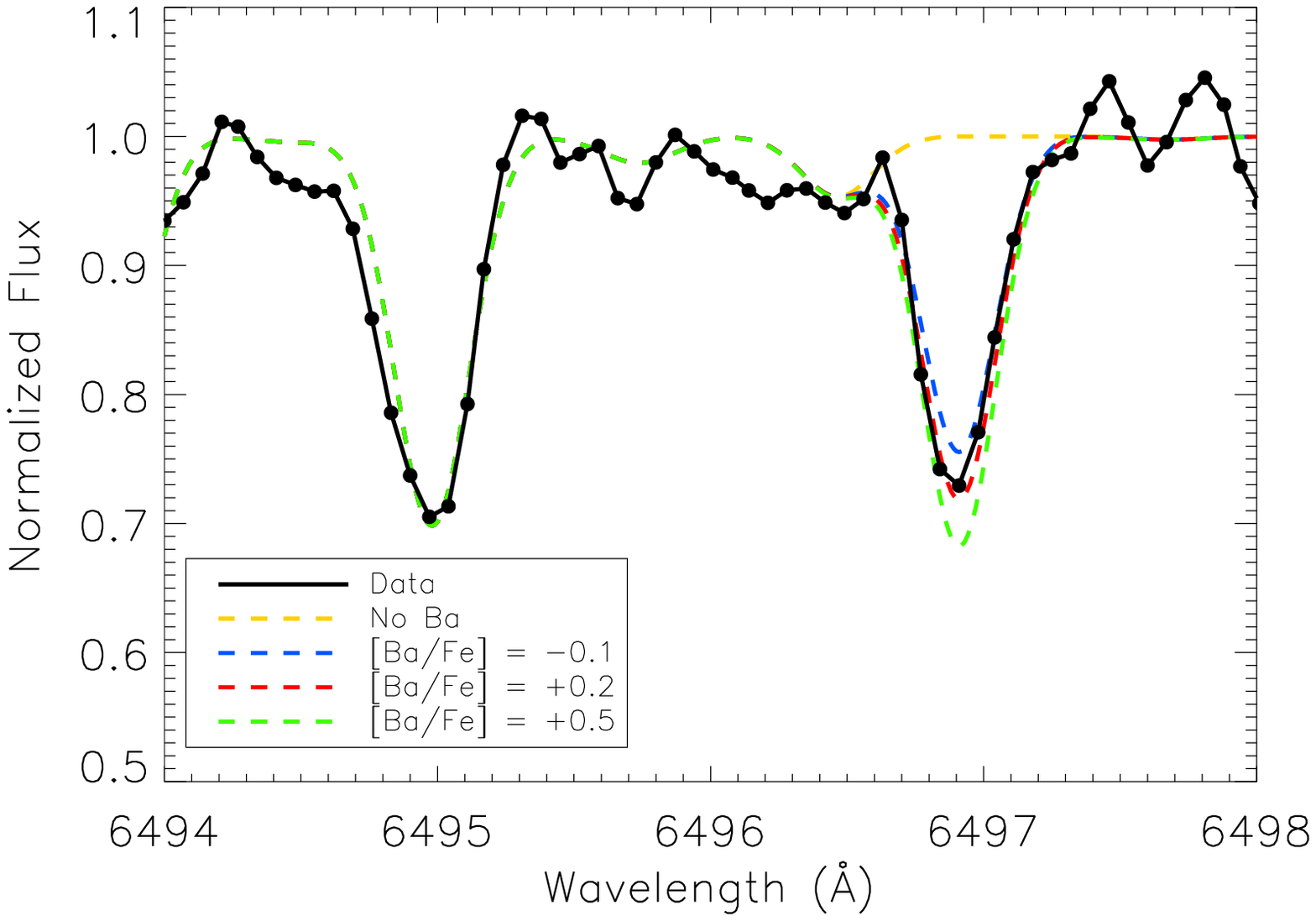} \\

  \includegraphics[width=8.5cm,clip=true,bbllx=80, bblly=362,bburx=558, bbury=705]{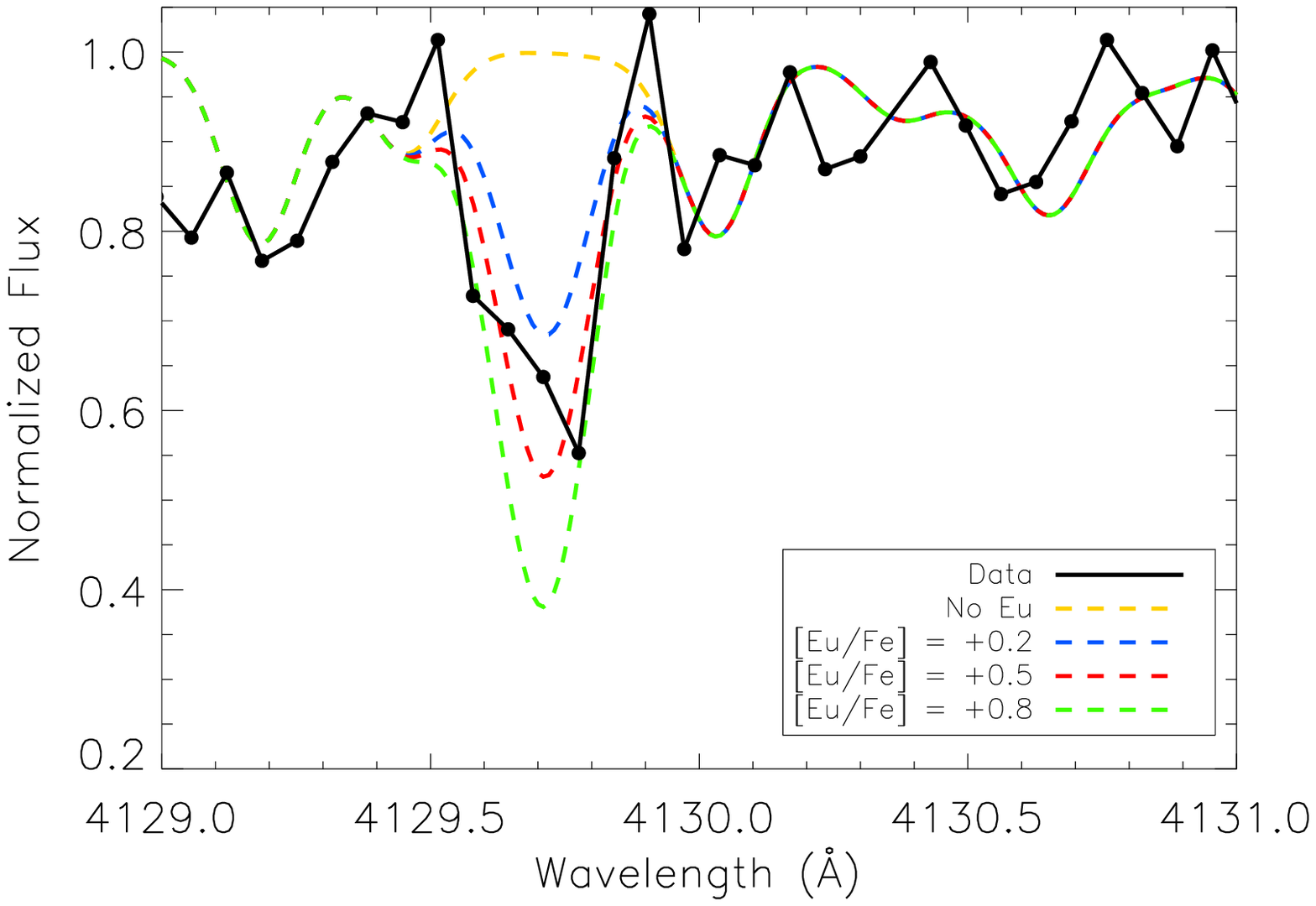} &  
  \includegraphics[width=8.5cm,clip=true,bbllx=80, bblly=362,bburx=550, bbury=705]{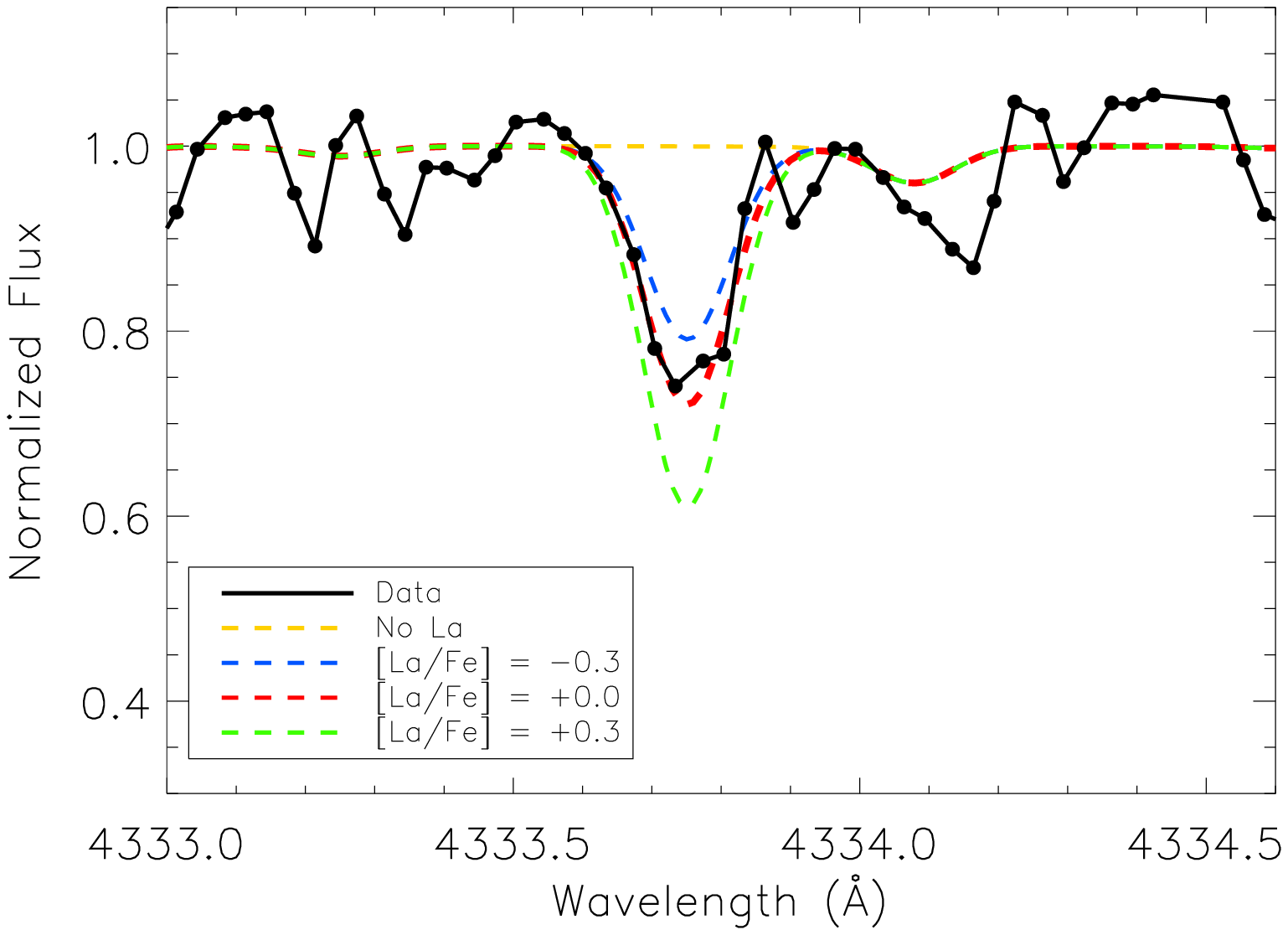} \\
  \end{tabular}
  \caption{Determining the abundances of Sr, Ba, Eu and La by
    comparing the observed lines to synthesized spectra at different
    abundances. }
  \label{fig:nc_syn}
 \end{center}
\end{figure*}

Strontium abundance was determined by synthesis of the line at
4215\,{\AA}, illustrated in Figure~\ref{fig:nc_syn}. The line at
4077\,{\AA} is also visible in the spectrum but too noisy to use for
abundance determination; the data are however not inconsistent with
what is determined from the line at 4215\,{\AA}, within the
uncertainties. Given the noise level even at 4215\,{\AA}, compared to
the difference between the synthesized spectra, this value should be
regarded as uncertain. In particular, even though the Sr abundance
appears abnormally low compared to other stellar populations in
Figure~\ref{fig:nc_el}, this is likely not significant given the
uncertainty.

Barium abundance was determined by synthesis of the lines at 4554 and
6496\,{\AA}, with the abundance quoted in Table~\ref{tab:seg11_abund}
being the average of the two. The synthesis of the 6496\,{\AA}, line
is shown in Figure~\ref{fig:nc_syn}. We adopt an uncertainty of $\pm$
0.3\,dex. Europium was determined by synthesis of the line at
4129\,{\AA}. Like strontium, there is considerable uncertainty due to
the noisy spectrum. Lanthanum was determined by synthesis of the line
at 4333\,{\AA}. Other lines are visible but too noisy for more precise
abundance determination; the upper limits derived are however
consistent with the result derived from the line at
4333\,{\AA}. Synthesis of the 4333\,{\AA}, line is shown in
Figure~\ref{fig:nc_syn}.

\subsection{Uncertainties}
\label{sec:unc}
Random errors come from uncertainties in placing the continuum level;
we estimate the random uncertainty in the abundance of an element as
the standard error of the mean abundance determined from individual
lines. For elements that were determined from fitting just one line
with a synthetic spectrum (Sr, Eu and La), the error quoted in the
second column is the estimated fitting uncertainty.

Systematic errors arise from uncertainties in the stellar parameters,
as described in Section~\ref{sec:stpar}. To quantify this effect, we
repeated the analysis with the stellar parameters of 300S-1 changed by
$+150$ K, $+0.4$\,dex, and $+0.3$\,km\,s$^{-1}$ in temperature, log g
and microturbulence respectively, and record the corresponding change
in abundance. Table~\ref{tab:seg11_sigma} shows the result. The total
uncertainty is obtained by summing the individual components in
quadrature.

Another assessment of the uncertainties, given the modest data
quality, comes from comparing our abundances from the low $S/N$
spectra of the comparison stars with those of F00.
Table~\ref{tab:hipabund} shows the derived abundances of the
comparison stars, and lists the published values from F00 also. Our
abundances are in good agreement within the uncertainties, especially
when taking into account the different stellar parameter solution for
HIP37335.

\begin{deluxetable}{lrrrrr}
\tabletypesize{\scriptsize}
\tablecaption{Abundance Uncertainties for 300S-1 \label{tab:seg11_sigma}}
\tablehead{
  \colhead{Elem.} &
  \colhead{Random} &
  \colhead{$\Delta T_{eff}$} &
  \colhead{$\Delta \log g$} &
  \colhead{$\Delta v_{micr}$} &
  \colhead{Total} \\
  \colhead{}&
  \colhead{Uncer.}&
  \colhead{+150\,K}&
  \colhead{+0.4\,dex}&
  \colhead{+0.3\,km\,s$^{-1}$}&
  \colhead{Uncer.}}
\startdata
  C (CH)   &  0.21       &  0.30            & $-0.05$         & $-0.02$           & 0.37        \\
  Na\,I    &  0.08 	 &  0.16            & $-0.10$	      &	$-0.04$		  & 0.21	\\
  Mg\,I    &  0.07 	 &  0.16            & $-0.11$	      &	$-0.05$           & 0.21	\\
  Ca\,I    &  0.06 	 &  0.13	    & $-0.05$	      &	$-0.10$		  & 0.18	\\
  Sc\,II   &  0.05 	 &  0.02	    & $+0.16$	      &	$-0.02$		  & 0.17	\\
  Ti\,I    &  0.07 	 &  0.20	    & $-0.02$	      &	$-0.09$		  & 0.23	\\
  Ti\,II   &  0.06 	 &  0.03	    & $+0.15$	      &	$-0.09$		  & 0.19	\\
  Cr\,I    &  0.05 	 &  0.19	    & $-0.01$	      &	$-0.08$		  & 0.21	\\
  Mn\,I    &  0.07 	 &  0.15	    & $-0.03$	      &	$-0.10$		  & 0.20	\\ 
  Fe\,I    &  0.05 	 &  0.18	    & $-0.04$	      &	$-0.13$		  & 0.23	\\
  Fe\,II   &  0.05 	 &$-0.02$	    & $+0.16$	      &	$-0.09$		  & 0.19	\\
  Co\,I    &  0.12 	 &  0.25	    & $+0.00$	      &	$-0.20$		  & 0.34	\\ 
  Ni\,I    &  0.06 	 &  0.14	    & $+0.00$	      &	$-0.06$		  & 0.16	\\
  Zn\,I    &  0.25 	 &  0.04	    & $+0.10$	      &	$-0.05$		  & 0.28	\\
  Sr\,II   &  0.40 	 &  0.10	    & $+0.02$	      &	$-0.10$		  & 0.42	\\ 
  Ba\,II   &  0.21 	 &  0.09	    & $+0.08$	      &	$-0.18$		  & 0.30	\\
  La\,II   &  0.30       &  0.05            & $+0.15$         & $-0.05$           & 0.34        \\
  Eu\,II   &  0.40 	 &  0.05	    & $+0.15$	      &	$-0.05$		  & 0.43	\\ \hline
\enddata
\end{deluxetable}

\begin{deluxetable}{lrccccccc}
\tabletypesize{\scriptsize}
\tablecaption{Standard Star Abundances \label{tab:hipabund}}
\tablehead{
  \colhead{Element} &
  \colhead{$\log\epsilon (\mbox{X})$} &
  \colhead{$\sigma$} &
  \colhead{N} &
  \colhead{[X/H]} &
  \colhead{[X/Fe]}& 
  \colhead{$[\mbox{X/Fe}]_{\mbox{\tiny{F00}}}$} \\
  & (dex) & (dex) & & (dex) & (dex) & (dex)
}
\startdata
\\
       &          &       &HIP37335&              &             &             \\ \hline
C (CH) & 7.71   & 0.21    &  2    &    $ -0.72 $  &  $ +0.28 $  &   \nodata   \\
Na\,I  & 5.46   & 0.07    &  4    &    $ -0.78 $  &  $ +0.22 $  &   $+0.32 $  \\       
Mg\,I  & 7.05   & 0.12    &  4    &    $ -0.55 $  &  $ +0.45 $  &   $+0.63 $  \\       
Ca\,I  & 5.86   & 0.05    &  16   &    $ -0.48 $  &  $ +0.52 $  &   $+0.44 $  \\       
Sc\,II & 2.40   & 0.05    &  9    &    $ -0.75 $  &  $ +0.25 $  &   \nodata   \\       
Ti\,I  & 4.20   & 0.05    &  28   &    $ -0.75 $  &  $ +0.25 $  &   $+0.26 $  \\       
Ti\,II & 4.20   & 0.05    &  16   &    $ -0.75 $  &  $ +0.25 $  &   \nodata   \\       
Cr\,I  & 4.65   & 0.05    &  14   &    $ -0.99 $  &  $ +0.01 $  &   $-0.05 $  \\       
Mn\,I  & 4.61   & 0.05    &  3    &    $ -0.82 $  &  $ +0.18 $  &   \nodata   \\       
Fe\,I  & 6.50   & 0.05    &  124  &    $ -1.00 $  &  $  0.00 $  &  $(-1.26)$  \\       
Fe\,II & 6.48   & 0.05    &  15   &    $ -1.02 $  &  $ -0.02 $  &   \nodata   \\  
Co\,I  & 3.55   & 0.05    &  3    &    $ -1.44 $  &  $ -0.44 $  &   \nodata   \\
Ni\,I  & 5.30   & 0.05    &  24   &    $ -0.92 $  &  $ +0.08 $  &   $+0.10 $  \\   
Zn\,I  & 3.78   & 0.08    &   2   &    $ -0.78 $  &  $ +0.22 $  &   \nodata   \\       
Ba\,II & 1.43   & 0.21    &  2    &    $ -0.75 $  &  $ +0.25 $  &   $-0.02 $  \\
La\,II & 0.42   & 0.30    &  1    &    $ -0.68 $  &  $ +0.32 $  &   \nodata   \\
Eu\,II &$-0.19$ & 0.30    &  1    &    $ -0.71 $  &  $ +0.29 $  &   $+0.38 $  \\
\\      
       &          &       &HIP68807&                 &              &               \\ \hline
C (CH) & 6.76 &  0.21     &    2   &      $ -1.67 $  &   $ +0.19 $  &  \nodata      \\
Na\,I  & 4.36 &  0.05     &    3   &      $ -1.88 $  &   $ -0.02 $  &  $  -0.13  $  \\       
Mg\,I  & 6.17 &  0.05     &    6   &      $ -1.43 $  &   $ +0.43 $  &  $  +0.49  $  \\       
Ca\,I  & 4.89 &  0.05     &    17  &      $ -1.45 $  &   $ +0.41 $  &  $  +0.37  $  \\       
Sc\,II & 1.45 &  0.06     &    6   &      $ -1.70 $  &   $ +0.16 $  &  \nodata      \\       
Ti\,I  & 3.25 &  0.05     &    23  &      $ -1.70 $  &   $ +0.16 $  &  $  +0.20  $  \\       
Ti\,II & 3.45 &  0.05     &    18  &      $ -1.50 $  &   $ +0.36 $  &  \nodata      \\       
Cr\,I  & 3.61 &  0.05     &    14  &      $ -2.03 $  &   $ -0.17 $  &  $  -0.12  $  \\       
Mn\,I  & 3.40 &  0.07     &    3   &      $ -2.03 $  &   $ -0.17 $  &  \nodata      \\       
Fe\,I  & 5.64 &  0.05     &    127 &      $ -1.86 $  &   $  0.00 $  &  $ (-1.83) $  \\       
Fe\,II & 5.68 &  0.05     &    14  &      $ -1.82 $  &   $  0.04 $  &  \nodata      \\       
Ni\,I  & 4.48 &  0.05     &    12  &      $ -1.74 $  &   $ +0.12 $  &  $  -0.03  $  \\    
Zn\,I  & 2.84 &  0.20     &    1   &      $ -1.72 $  &   $ +0.14 $  &  \nodata      \\      
Ba\,II & 0.69 &  0.14     &    3   &      $ -1.49 $  &   $ +0.37 $  &  $  +0.27  $  \\
La\,II &$-0.68$& 0.30     &    1   &      $ -1.78 $  &   $ +0.08 $  &  \nodata      \\
Eu\,II &$-0.99$& 0.40     &    1   &      $ -1.51 $  &   $ +0.35 $  &  $  +0.40  $  \\
\\
       &          &         &HIP47139&               &             &                \\ \hline

C (CH) & 6.61     &  0.21   &   2    &    $ -1.82 $  &  $ -0.23 $  &  \nodata       \\
O\,I   & 8.14     &  0.15   &   1    &    $ -0.55 $  &  $+ 1.04 $  &  \nodata       \\
Na\,I  & 4.65     &  0.05   &   4    &    $ -1.59 $  &  $  0.00 $  &  $ -0.18   $   \\       
Mg\,I  & 6.42     &  0.08   &   4    &    $ -1.18 $  &  $ +0.41 $  &  $ +0.54   $   \\       
Ca\,I  & 5.05     &  0.05   &   16   &    $ -1.29 $  &  $ +0.30 $  &  $ +0.27   $   \\       
Sc\,II & 1.64     &  0.05   &   12   &    $ -1.51 $  &  $ +0.08 $  &  \nodata       \\       
Ti\,I  & 3.52     &  0.05   &   26   &    $ -1.43 $  &  $ +0.16 $  &  $ +0.29   $   \\       
Ti\,II & 3.67     &  0.05   &   18   &    $ -1.28 $  &  $ +0.31 $  &  \nodata       \\       
Cr\,I  & 3.94     &  0.07   &   16   &    $ -1.70 $  &  $ -0.11 $  &  $ -0.17   $   \\       
Mn\,I  & 3.58     &  0.07   &   3    &    $ -1.85 $  &  $ -0.26 $  &  \nodata       \\       
Fe\,I  & 5.91     &  0.05   &   109  &    $ -1.59 $  &  $  0.00 $  &  $ (-1.46) $   \\       
Fe\,II & 5.94     &  0.05   &   11   &    $ -1.56 $  &  $  0.03 $  &  \nodata       \\       
Ni\,I  & 4.65     &  0.05   &   18   &    $ -1.57 $  &  $ +0.02 $  &  $ +0.00   $   \\  
Zn\,I  & 2.97     &  0.11   &   2    &    $ -1.59 $  &  $ -0.00 $  &  \nodata       \\     
Ba\,II & 0.92     &  0.10   &   3    &    $ -1.26 $  &  $ +0.33 $  &  $ +0.16   $   \\
La\,II & $-0.48$  &  0.30   &   1    &    $ -1.58 $  &  $ +0.01 $  &  \nodata       \\
Eu\,II & $-0.49$  &  0.40   &   1    &    $ -1.01 $  &  $ +0.58 $  &  \nodata       
\enddata
\end{deluxetable}


\section{Characterizing the 300\,km\,s$^{-1}$ Stream}
\label{sec:char}

\subsection{Stream Membership}
\begin{figure*}
 \includegraphics[width=\columnwidth]{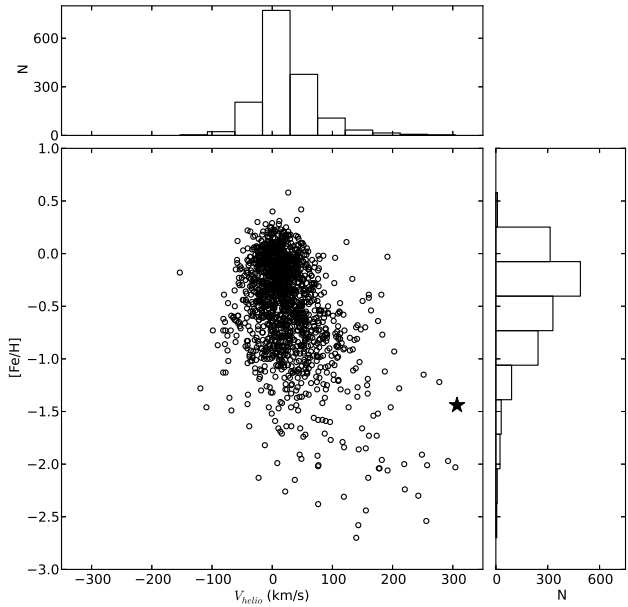}
 \caption{Kinematics and metallicities for line-of-sight stars
   predicted by the Besan\c{c}on model \citep{Robin2003}. The star
   analysed here, 300S-1, is marked as a filled star.}
 \label{fig:besancon}
\end{figure*}

As demonstrated by other authors, there is unequivocally a coherent
stream present here with a kinematic peak at $v_{helio} = 300$ km
s$^{-1}$ \citep{Geha2009, Norris2010a, Simon2011}. With more extreme
velocities, halo contaminants become less likely. Given that a stream
is present here with high velocities, it is worth quantifying the
probability whether the star analysed here is a background halo star
or not. We have employed a two-sample Kolmogorov-Smirnov test using
the predicted line-of-sight velocities from the Besan\c{c}on model
\citep{Robin2003} to quantify this likelihood. We find a $p$ value of
$0.097$, or a $\sim10$\% chance that 300S-1 and predicted stars in the
Besan\c{c}on model \citep{Robin2003} are drawn from the same
distribution. It is clear that this star sits separate from the main
line-of-sight predicted population (Figure \ref{fig:besancon}).

We can also deduce some likelihood that 300S-1 is a halo member when
we examine the observed velocity distribution in Figure
\ref{fig:col_mag}. The peak of the stream velocity distribution occurs
at $300$\,km s$^{-1}$ and comprises 39 member stars, amongst
background halo outliers with velocities between $250$ and $400$\,km
s$^{-1}$. Within the $270-330$\,km s$^{-1}$ range, it is reasonable to
suspect that we would observe fewer than 10 halo contaminants in
$2^{\circ}$ with such high velocities. Indeed, the observed background
range in Figure \ref{fig:col_mag} is approximately 5-10 halo stars,
suggesting a probability of between $13-26\%$ that 300S-1 is a halo
contaminant. We note that the probability that this star is a halo
star ($\gtrsim10$\%) is the same probability cutoff employed by
\citet{Simon2011} in examining stream members.

It is well-known that RGB stars in globular clusters exhibit a
characteristic anti-correlation in Na-O, and Al-Mg
\citep{Carretta2009}. Due to the low spatial distribution and
relatively high kinematic dispersion for this stream compared to
typical kinematically cold stellar streams, it is possible that the
origin of the 300\,km\,s$^{-1}$ stream is a disrupted globular
cluster. Although our spectrum has extremely modest S/N below
4000\,{\AA}, we have attempted to synthesize the Al lines at
3944\,{\AA} and 3961\,{\AA}. The Al lines at $\sim6697$\,{\AA} were
not detected. Hence, we cannot determine an accurate value
for [Al/Fe] from the blue lines. We can only exclude a super-solar
abundance $\mbox{[Al/Fe]}<0$ for this star. However, this is a rather
low [Al/Fe] abundance if 300S-1 was a member of a globular
cluster. While there are globular cluster stars with
$\mbox{[Al/Fe]}<0$ and similar metallicities, they generally have Mg
abundances of $0.3 < \mbox{[Mg/Fe]} <0.6$ (compare to
\citealt{Carretta2009}, their Figure 5). 300S-1 has $\mbox{[Mg/Fe]} =0.14$
and even if that Mg abundance would be systematically low by 0.1 to
0.2\,dex, it would still mostly fall outside any covered
region. It suggests that 300S-1 may not be of a globular cluster
origin.  Unfortunately, given the modest S/N of our spectra no
reliable upper limit on oxygen could be ascertained that could provide
further clues on the topic.

\begin{figure*}
 \begin{center}
  \includegraphics[width=17cm,clip=true,bbllx=40, bblly=362,bburx=555, bbury=705]{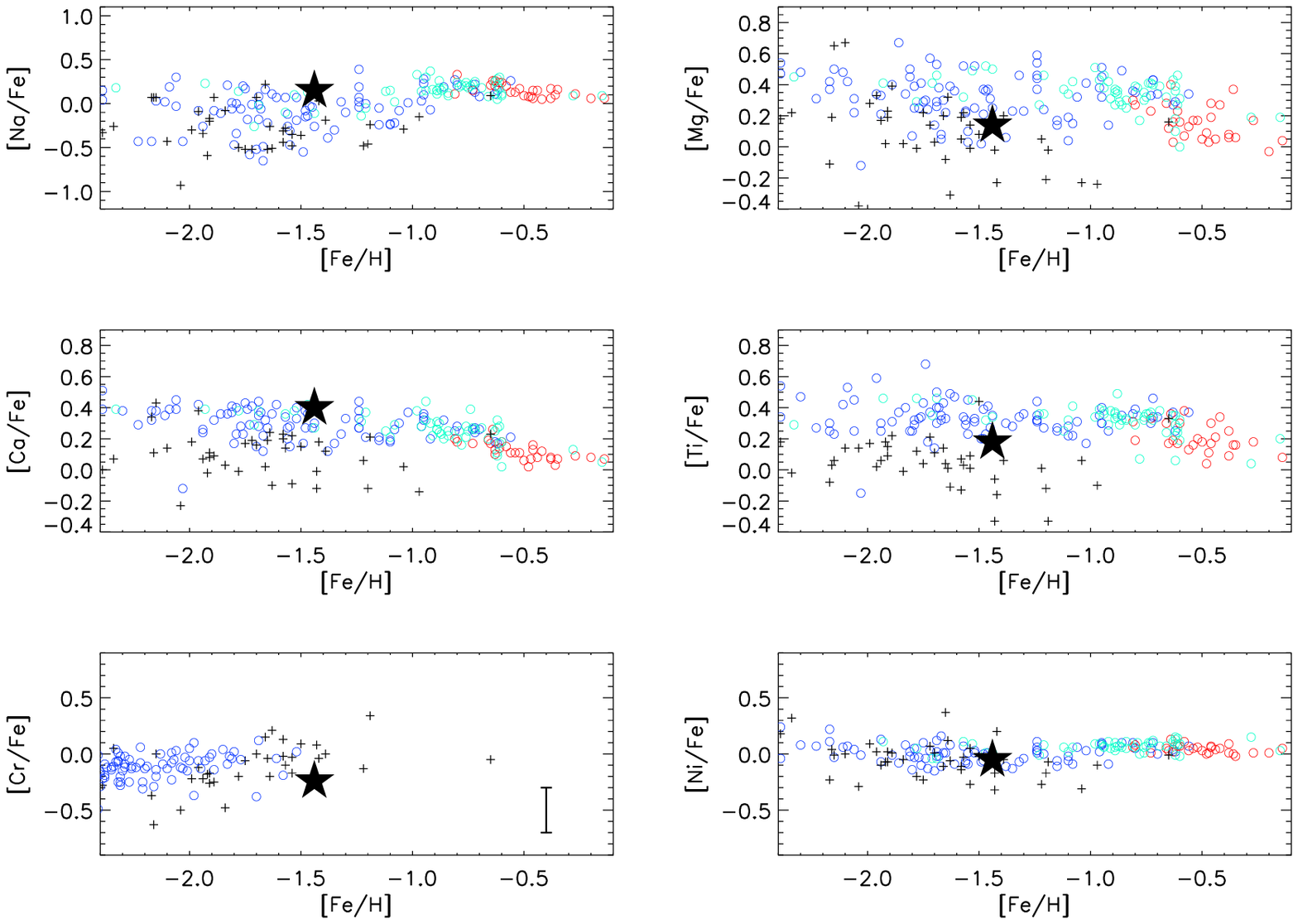}
  \caption{Abundance ratios for light elements in 300S-1 (filled
    star symbol), as measured from our high-resolution spectrum. For
    comparison, red and cyan points show thin and thick disk stars
    respectively, blue points show halo stars, and the crosses show
    stars in the classical dwarf galaxies Draco, Sextans, Ursa Minor,
    Carina, Fornax, Sculptor and Leo I. A typical error bar for our
    measurements is shown in the lower-left panel; also see
    Table~\ref{tab:seg11_sigma}. }
  \label{fig:light_el}
 \end{center}
\end{figure*}

\begin{figure*}
 \begin{center}
  \includegraphics[width=17cm,clip=true,bbllx=80, bblly=362,bburx=555, bbury=705]{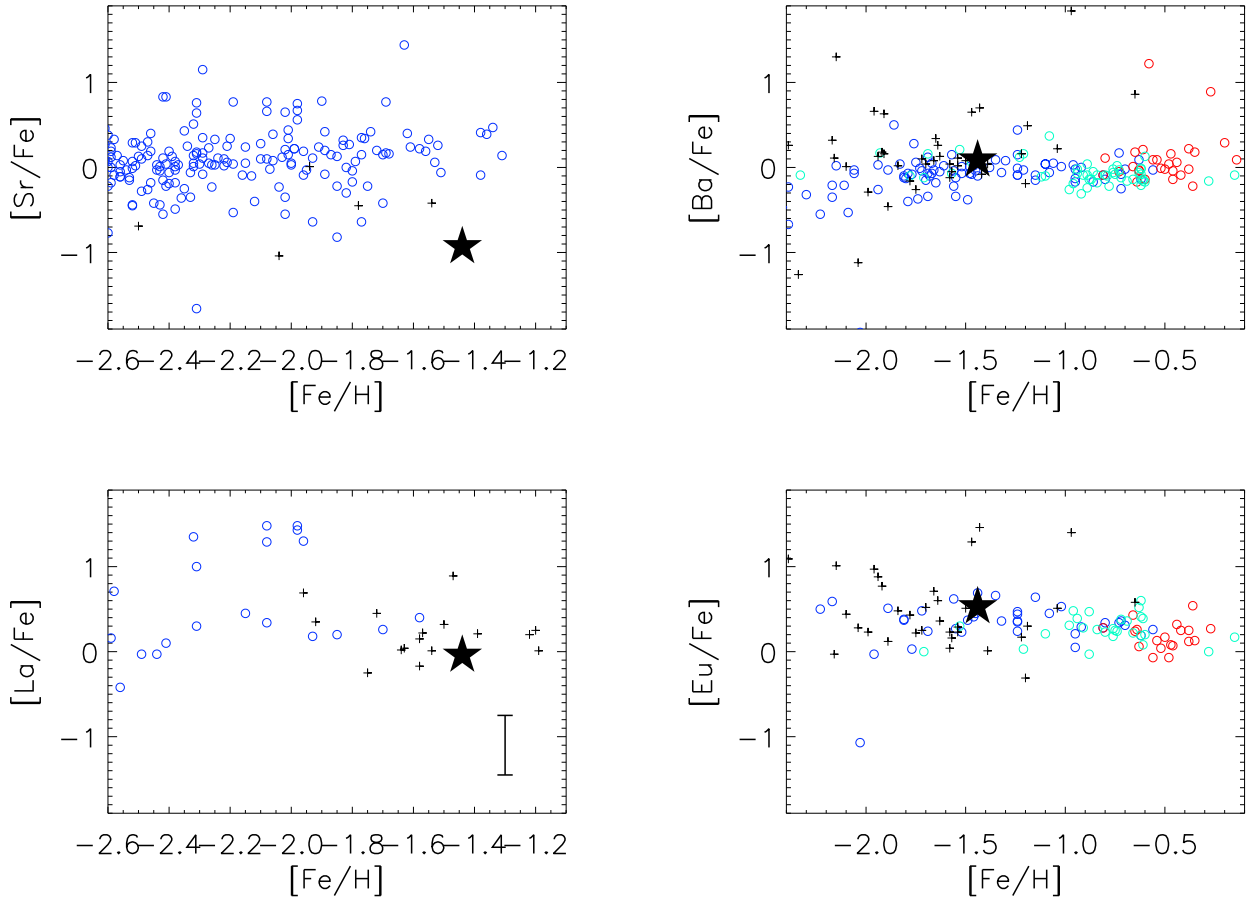}
  \caption{Abundance ratios for the neutron-capture elements Sr, Ba,
    La and Eu in 300S-1. Symbols as in Figure~\ref{fig:light_el}.}
  \label{fig:nc_el}
 \end{center}
\end{figure*}

\subsection{Metallicity of Stream}
We determine a metallicity of $\mbox{[Fe/H]} = -1.46 \pm 0.05 \pm 0.23
$ (random and systematic uncertainties) for 300S-1 based on the
high-resolution spectrum. This is in agreement with the prediction of
$\mbox{[Fe/H]} = -1.3$ from \citet{Simon2011}. This is also consistent
with what we roughly estimate from fitting the M5 (with
$\mbox{[Fe/H]}=-1.2$) isochrone to the stream photometry.  It adds
support to the result that the stream stars have higher metallicity
than the Segue\,1 system, as already noted in S11 based on Ca triplet
equivalents widths. The AAOmega sample of N10 contains some
metallicity estimates based on Ca\,II K line strengths; in addition to
300S-1, the spectrum of the stream star Segue1-101
(SDSSJ100659.01+154418.8) has enough counts to yield $\mbox{[Fe/H]}
\simeq -1.7$. Beyond this, there is no additional data to determine
the metallicity spread of the stream.

\subsection{Evolutionary Status}

From the spectroscopic analysis we find 300S-1 to be a red giant
branch star. Although the star sits slightly above the isochrone, it
agrees with the isochrone within the uncertainties in the stellar
parameters.  From photometry, the star is also found to sit slightly
above a shifted M5 globular cluster sequence. It is noteworthy,
though, that the scatter of stream members is significant (see
Figure~1). Nevertheless, these discrepancies could indicate the star
to be on the horizontal branch. The carbon abundance of 300S-1 may
shed light on this question. As a star ascends the giant branch, CN
cycling converts carbon into nitrogen, thus lowering the observed
surface carbon abundances. The measured value of $\mbox{[C/Fe]}=0.25$
suggests that CN cycling has not yet significantly operated, assuming
that the star did not form from an unusually carbon-rich gas
cloud. Some globular clusters show this effect (e.g.,
\citealt{carretta05}, their Figure~7) and compared to that, the carbon
abundance of 300S-1 also suggests the star to be on the lower or
middle part of the red giant branch and not in a more evolved state on
the horizontal branch.

\subsection{Abundance Ratios}
\label{sec:ab_rat}
Figures~\ref{fig:light_el} and \ref{fig:nc_el} show the abundance
patterns of 300S-1, based on our high-resolution spectrum. The
blue, cyan and red points show stars in the halo, thick disk and thin
disk respectively, from the sample of \citet{Fulbright2000,
  Fulbright2002}. Since the
Fulbright sample does not include Cr, Sr and La measurements,
comparison points (halo stars) for these elements are taken from
\citet{Lai2007} and \citet{Barklem2005}. In addition, the crosses show
abundance patterns of stars in the classical dwarf galaxies Draco,
Sextans, Ursa Minor, Carina, Fornax, Sculptor and Leo I
\citep{Shetrone2001, Shetrone2003, Geisler2005, Aoki2009, Cohen2009}.

The abundance ratios of 300S-1 overlap well with the general halo
population at that metallicity, though as noted in
Section~\ref{sec:abund}, the $\alpha$- and particularly Mg abundances
are on the low end. (The Sr abundance also appears low, but as
discussed in Section~\ref{sec:abund}, this is likely not significant
due to the low $S/N$ of the spectrum in this region.) The general
abundance pattern is closer to a typical halo star than to a
classical dwarf spheroidal galaxy star.

The overall abundance pattern is important for interpreting the nature
of the stream and its potential progenitor. Judging from this one
star, 300S may be the first stream with halo-like abundances. This
predicament illustrates that a careful mapping of the region around
Segue\,1 and the 300\,km\,s$^{-1}$ stream is of great
importance. Initial work on SDSS data to address this question hints
that this region is even more complex than assumed thus far. An
attempt to decompose the many populations in the Segue\,1 region will
be presented in a forthcoming paper (A. Jayaraman et al., in
preparation).

\subsection{Distance to stream}

We use two different methods of estimating the distance to the
stream. First, based on the stellar parameters we assume 300S-1 to
be a red giant (Figure~\ref{fig:isochr}). Given the star's
metallicity, we chose the $\mbox{[Fe/H]} = -1.5$ isochrone as the ``best
fit''. While it is not a perfect match to our stellar parameters, it
is reasonably close to our determined values. The corresponding
inferred absolute magnitude of 300S-1 is $M_V \simeq
+1.37$. Transforming the Sloan photometry following
\citet{Jordi2006}, and correcting for extinction according to
\citet{Schlegel1998}, we find 300S-1 has $V = 17.60$. This yields a
distance modulus of $16.23$, resulting in a best distance estimate of
$\simeq 18$\,kpc. Given the uncertainties, however, a distance within
$\pm 7$\,kpc of this estimate would still be consistent with the
stellar parameters we derived from the spectroscopy. Our distance
estimate of 18\,kpc is in good agreement with \citet{Simon2011} who
find a distance of 22\,kpc to the stream.

A second distance estimate comes from using the photometric data and
comparing the color-magnitude diagram to various globular cluster
sequences from \citet{An2008}, shifted according to the distance
moduli and reddening ($E(B-V)=0.03$ for M5, $E(B-V)=0.02$ for M92)
compiled in \citet{Harris1996}. As seen in Figure~\ref{fig:col_mag},
the M5 sequence is a good fit to the stream data when shifted to a
distance of 18\,kpc. This is in good agreement with the estimate based
on spectroscopically determined stellar parameters.

Both the isochrone fit based on the single spectroscopic measurement
and the photometric data indicate that the stream stars are at a
distance of $\simeq 18$\,kpc, slightly closer than the assumed
distance of $23 \pm 2$\,kpc \citep{Belokurov2007} for Segue\,1
itself. However, the uncertainties on both values are substantial
enough that we cannot rule out that they are at the same distance. We
also caution that since the stream stars were picked out in
color-magnitude filters targeting stars at the distance of Segue\,1, the
stream stars in our sample by design cannot be at a very different
distance.

We note that with galactic coordinates $l, b \simeq 220^{\circ},
50^{\circ}$ and a heliocentric distance of 18\,kpc, the stream stars
are located in the outer galaxy. A heliocentric radial velocity of
300\,km\,s$^{-1}$ translates to Galactic Standard of Rest velocity of
about 230\,km\,s$^{-1}$ in this direction. This suggests the stream
stars to be on a low angular momentum orbit that would eventually
bring it closer and into the inner Galaxy.

\section{Conclusions} \label{sec:conc}

We have presented a high-resolution spectrum and abundance analysis of
300S-1, a bright star in the 300\,km\,s$^{-1}$ stream near the
ultra-faint dwarf galaxy Segue\,1. We determine a metallicity
$\mbox{[Fe/H]} = -1.46 \pm 0.05 \pm 0.23$ (random and systematic
uncertainties) for this star, with abundance ratios similar to typical
halo stars at this metallicity. Fitting the stellar parameter solution
onto theoretical isochrones, we estimate a distance of $18 \pm
7$\,kpc. Both the metallicity and distance are in good agreement with
estimates obtained from comparing the SDSS photometry to globular
cluster sequences.

With this new information, we present several possible scenarios
regarding the nature and origin of the stream:

\begin{itemize}
\item Since these high-velocity stars were discovered in a survey
  targeting Segue 1 members, a natural question to ask is whether the
  stream is related to the Segue\,1 dwarf galaxy. We find this an
  unlikely scenario for several reasons. First, the study of S11 finds
  no evidence that the Segue\,1 system is being tidally disrupted. Our
  distance estimate, based both on the high-resolution spectroscopy
  data and the photometry indicate that the stream stars are at a
  slightly closer distance than Segue\,1, though the data are not good
  enough to rule out that they are at the same distance. In addition,
  the color-magnitude diagrams suggest that the stream members are in
  general at a higher metallicity than Segue\,1, which our
  high-resolution measurement of one stream star also confirms.

\item Another possibility, suggested by \citet{Geha2009}, is that
  these stars could be associated with the Sagittarius stream. Indeed
  at least two wraps of Sagittarius overlap with Segue 1 in this
  direction, and \citet{Niederste-Ostholt2009} argue that Segue 1
  itself is a star cluster from the Sagittarius galaxy. But as for the
  stream stars, our metallicity measurement indicates that the
  300\,km\,s$^{-1}$ stream does not have metallicities representative
  of Sgr debris, \citep{Chou2007, Casey2012}, and no Sgr debris model
  that we are aware of predicts a wrap at this velocity. Moreover, the
  part of the Sagittarius stream proposed to be contaminating Segue 1
  samples has $v_{GSR} \sim 130$\,km\,s$^{-1}$
  \citep{Niederste-Ostholt2009}, while the stream stars have $v_{GSR}
  \sim 230$\,km\,s$^{-1}$.

\item The Orphan Stream \citep{Belokurov2007b} crosses the Sagittarius
  stream on the sky near Segue\,1, and at a similar distance
  modulus. Again, however, the velocities do not agree -- the reported
  velocities of the Orphan Stream are around $v_{GSR} \sim
  110$\,km\,s$^{-1}$. In addition, the Orphan stream is metal-poor,
  with reported metallicities of $\mbox{[Fe/H]} = -1.63$ to $-2.10$
  \citep{Newberg2010,casey13}. Although both the Orphan Stream and
  Sagittarius Stream overlap with the 300\,km\,s$^{-1}$ stars, then,
  the combined metallicity and velocity information suggest that they
  are unrelated.

\item We have assumed throughout this paper that the 300\,km\,s$^{-1}$
  feature is a stellar stream, but as pointed out in S11, until we can
  determine the full spatial extent of these kinematically linked
  stars, we cannot rule out that the stars belong to a bound
  object. If so, given the 1$^{\circ}$ extent seen in N10, its
  physical diameter would be at least 300 ($d / 18$\,kpc) pc. This
  highlights the need for new photometry to map out the full
  extent of the 300\,km\,s$^{-1}$ stars.

\end{itemize}

The fact that this stream may be largely chemically similar to the
halo is particularly interesting. This is relevant to chemical tagging
\citep{Freeman_Bland-Hawthorn2003}, which infers that stars
originating from a common origin can be unambiguously identified
solely by their chemistry and without the need for kinematics. This
stream is most noticable only from its kinematics, and not by any
particularly distinct chemical signature identified in this
work. Although the luminosity, number of abundances analyzed or
abundance uncertainties presented here do not match the strict
requirements for the complete chemical tagging planned in future
galactic archaeology surveys \citep{Ting2012}, we identify this
300\,km\,s$^{-1}$ stream as a candidate for testing and validating the
chemical tagging concept. It would be particularly interesting to
determine whether chemical tagging alone could identify members
belonging to this 300\,km\,s$^{-1}$ stream without the need for
kinematics, as the chemical elements analyzed here are only marginally
distinguishable from the halo.

Although the 300\,km\,s$^{-1}$ stars are found in a region of sky with
many known structures, the combination of velocity, chemistry and
distance information makes it unlikely that these stars are associated
with any of the Sagittarius stream, the Orphan stream, or the Segue\,1
dwarf galaxy. We therefore conclude that these stars belong to a new
structure in the crowded ``Field of Streams''. Its features include an
extreme mean velocity of 300\,km\,s$^{-1}$ with a velocity dispersion
of 7\,km\,s$^{-1}$ (as found by S11), a broad spatial distribution,
and halo-like chemical abundances. The abundance patterns in
particular make this stream very interesting to study in the context
of halo formation.

\acknowledgements{ A.F. acknowledges support of an earlier Clay
  Fellowship administered by the Smithsonian Astrophysical
  Observatory. A.R.C. acknowledges the financial support through the
  Australian Research Council Laureate Fellowship 0992131, and from
  the Australian Prime Minister's Endeavour Award Research Fellowship,
  which has facilitated his research at MIT. J.E.N. acknowledges
  support from the Australian Research Council (grants DP063563 and
  DP0984924) for studies of the Galaxy's most metal-poor stars and
  ultra-faint satellite systems. R.F.G.W. acknowledges support from
  NSF grants AST-0908326 and CDI-1124403.}

\textit{Facilities:} \facility{Magellan-Clay (MIKE)}

\end{document}